\documentclass[useAMS,usenatbib]{mn2e}

\usepackage{graphicx,amssymb,bm}
\usepackage{subfig,xfrac}
\usepackage{float, Abbrev}
\usepackage[fleqn]{amsmath}  
\usepackage{color}
\usepackage[draft]{hyperref}

\usepackage{multirow}

\newcommand{\vect}[1]{\boldsymbol{#1}}

\newcommand{\be}{\begin{equation}}
\newcommand{\ee}{\end{equation}}
\newcommand{\ba}{\begin{eqnarray}}
\newcommand{\ea}{\end{eqnarray}}

\def\simlt{\lower.5ex\hbox{$\; \buildrel < \over \sim \;$}}

\newcommand{\fig}{\begin{figure} \begin{center}}
\newcommand{\efig}{\end{center}\end{figure} }
\newcommand{\figs}{\begin{figure*}\begin{minipage}{180mm} \begin{center}}
\newcommand{\efigs}{\end{center}\end{minipage}\end{figure*} }
\def\simgt{\lower.5ex\hbox{$\; \buildrel > \over \sim \;$}}

\usepackage{graphicx} 

\title[The shape of clusters]{Reconciling galaxy cluster shapes, measured by theorists vs observers}
\author[D. Harvey et al]
{David Harvey$^{1}$\thanks{e-mail: {\tt harvey@lorentz.leidenuniv.nl}}, {Andrew Robertson}$^{2,3}$, {Sut-Ieng Tam}$^{2}$, {Mathilde Jauzac}$^{2,3,4}$, \newauthor {Richard Massey}$^{2,3}$,
 {Jason Rhodes}$^{5,6}$,  and {Ian G. McCarthy}$^{7}$ \\
$^{1}$Lorentz Institute, Leiden University, Niels Bohrweg 2, Leiden, NL-2333 CA, The Netherlands \\
$^{2}$Institute for Computational Cosmology, Durham University, South Road, Durham DH1 3LE, UK\\
$^{3}$Centre for Extragalactic Astronomy, Department of Physics, Durham University, Durham DH1 3LE, U.K. \\
$^{4}$Astrophysics and Cosmology Research Unit, School of Mathematical Sciences, University of KwaZulu-Natal, Durban 4041, South Africa\\
$^{5}$Jet Propulsion Laboratory, California Institute of Technology, Pasadena, CA 91109, USA\\
$^{6}$California Institute of Technology, 1201 East California Blvd., Pasadena, CA 91125, USA\\
$^{7}$Astrophysics Research Institute, Liverpool John Moores University, 146 Brownlow Hill, Liverpool, L3 5RF, UK}
\begin{document}

\date{Accepted ---. Received ---; in original form \today.}

\pagerange{\pageref{firstpage}--\pageref{lastpage}} \pubyear{2017}

\maketitle

\label{firstpage}

\begin{abstract}
If properly calibrated, the shapes of galaxy clusters can be used to investigate many physical processes: from feedback and quenching of star formation, to the nature of dark matter. Theorists frequently measure shapes using moments of inertia of simulated particles'. We instead create mock (optical, X-ray, strong- and weak-lensing) observations of the twenty-two most massive ($\sim10^{14.7}\,M_\odot$) relaxed clusters in the BAHAMAS simulations. We find that observable measures of shape are rounder. Even when moments of inertia are projected into 2D and evaluated at matched radius, they overestimate ellipticity by 56\% (compared to observable strong lensing) and 430\% (compared to observable weak lensing). Therefore, we propose matchable quantities and test them using observations of eight relaxed clusters from the {\emph Hubble Space Telescope} and {\emph Chandra X-Ray Observatory}. We also release our HST data reduction and lensing analysis software to the community. In real clusters, the ellipticity and orientation angle at all radii are strongly correlated. In simulated clusters, the ellipticity of inner ($<r_{\mathrm{vir}}/20$) regions becomes decoupled: for example with greater misalignment of the central cluster galaxy. This may indicate overly efficient implementation of feedback from active galactic nuclei. Future exploitation of cluster shapes as a function of radii will require better understanding of core baryonic processes. Exploitation of shapes on any scale will require calibration on simulations extended all the way to mock observations. 

\end{abstract}


\section{Introduction}
The $\Lambda$CDM concordant model of cosmology assumes that we are living in a Universe dominated by an unknown dark energy, accelerating the expansion of space-time and permeated by a dominant gravitating mass that we do not understand, dark matter \citep{planckpars,chftpars,wiggles,DEScosmology,kids450}. In this framework, cosmological simulations predict that structure should form hierarchially, with the smallest haloes forming first \citep{Illustris,springel01,sdssboss,BAHAMAS,eagle,illustrisTNG}. The resulting distribution of matter is in the form of a ``cosmic web'' whereby filaments and sheets of mass funnel mass towards massive nodes or galaxy clusters.

Galaxy clusters are the largest known virialised structures in the Universe. They are dominated by dark matter, harbouring a halo of hot X-ray gas and in some cases thousands of galaxies \citep[e.g][]{Locuss_Smith,CLASH,HFF}. As extreme peaks in the density field, clusters of galaxies are ideal laboratories to study dark matter \citep[e.g.][]{Harvey_trails,Harvey_BCG,substructure_a2744A,substructure_a2744_wavelets,SIDM_shapes_subhalo,SIDMModel,RobertsonBAHAMAS,SIDM_BAHAMAS} and constrain cosmology \citep[e.g.][]{peakCosmology,peaksnongauss,peakmodifiedGR}. 

The mass of a galaxy cluster can exceed $10^{15}$ solar masses \citep[e.g.][]{MACSJ0717_HFF_zitrin,WtG,CLASH_zitrin,Merten_clash,harvey_0416,cosmicBeast}. In these environments space-time is deformed causing the observed image of distant galaxies that happen to align themselves with the cluster to be distorted. In extreme cases, the images can be stretched into arcs and split into multiple copies. Modelling strong gravitational lensing has become common place when measuring the mass distribution in clusters of galaxies \citep[e.g][]{CLASH_zitrin,Merten_clash,MACSJ1149_HFF}. However, strong lensing has its limitations, with the constraints limited to the  core of the cluster, there is no information on substructures and mass in its outer regions. Weak gravitational lensing, where the effect of the cluster must be measured statistically, grants access to this missing information \citep[e.g.][]{A1758,WtG}. It is now normal to combine both weak and strong gravitational lensing to get a full picture of the cluster \citep[e.g.][]{CLASH_zitrin,Merten_clash,strongweakunited1,strongweakunited2,MERTEN}. For a full review of mass mapping in clusters of galaxies see \cite{massModelReview}.

The exact form of galaxy clusters is still debated however it is generally accepted that they are triaxial in their shape with X-ray studies suggesting that roughly 70\% of all clusters tend to be prolate \citep{xrayShapes}. With X-ray shapes prone to non-thermal support, studies cited gravitational lensing as a more direct method to measure the shape of clusters. \cite{lensingShapes} studied the weak lensing properties of 25 clusters of galaxies, finding an average ellipticity, $\langle \epsilon\rangle = 0.46\pm0.04$. They interestingly found no correlation between both the lensing ellipticity and position angle of the member galaxies, suggesting no connection.
More recently, \cite{clashMorph} carried out a study where they identified the connection between the X-ray emission, the Brightest Cluster Galaxy (BCG) and lensing (strong and weak combined). They found that there was a strong correlation between the position angle of the cluster at large radial distances ($r\sim500$kpc), and the BCG (inner $10$kpc), citing a coupling between the cluster and galactic star-formation properties. 
Finally a recent study of twenty relaxed and dynamically merging clusters looked at the mis-alignment of morphologies between the weak lensing and four probes, the Sunyaev-Zel'dovich effect, the X-ray morphology, the strong lensing morphologies and the BCG \citep{CLASHalignment}. 

Although many studies have measured the shapes of clusters, fewer  studies exist comparing the shapes to hydro-dynamical simulations \citep[e.g.][]{CLASHalignment}. In these studies it is normal to measure the inertia tensor directly from the particle data. By taking eigenvectors and eigenvalues it is possible to extract directly the axis ratios and position angles. However, the radii in which the particles are chosen to measure this inertia tensor greatly impacts the inferred shape, and as such it is not clear how this shape relates to the weak or strong lensing \citep{EagleIntrinsicAlign}.

In addition, sample matching is often overlooked. The majority of clusters that have been observed with deep enough imaging to measure strong-lensing have extremely complicated selection functions, and the clusters selected are often highly irregular. For example the Hubble Frontier Fields contains clusters that are incredibly complicated, with some studies even suggesting that they are in tension with $\Lambda$CDM \citep[e.g A2744,][]{substructure_a2744,substructure_a2744A}. As such, it raises the question of the comparability of the full sample of currently observed clusters with simulations. A recent study compared the complete sample of CLASH, HFF and RELICS with simulations from the Horizon-AGN suite of simulations \citep{clusterCompWithHorizon}. An interesting study that found the observed clusters appeared to have strong lensing regions significantly more elliptical than the BCG. However, many of the clusters used are massive merging clusters, (for example A2744, MACSJ0416, MACS1149.5), that have huge structures in-falling, which will bias the results should the simulations not reflect the identical sample.

In this paper we study two key questions:
\begin{enumerate}
\item Is the ellipticity calculated from the projected moment of inertia derived directly from particle data in simulations a good estimator of the shape derived from strong or weak lensing? 
\item Is there any evidence for a radial dependent ellipticity in galaxy clusters, potentially signalling different physics acting on different regions of the cluster?
\end{enumerate}
To answer these two questions we carry out our investigation with two key differences to previous studies. First, we derive observationally matched products from the simulations so we can carry out a parallel analysis and make a direct comparison to observations, and secondly, we impose a strict selection cut on both the observed and simulated sample of clusters to ensure that they are analogous.

The paper is structured as follows: In section \ref{sec:Data} we outline the data that is used, including the reduction processes used and the simulations that have been adopted. In Section \ref{sec:method} we briefly review how we measure the ellipiticity from different probes (with a full description in Appendix \ref{sec:shapes}.) In section \ref{sec:results} we show our results and then in section \ref{sec:conc} we conclude.

\section{Data}\label{sec:Data}

\subsection{N-body simulations of galaxy clusters}
We use the BAHAMAS suite of simulations to compare the shapes of the observed clusters to clusters in a $\Lambda$CDM universe \citep{BAHAMAS,BAHAMASB}. BAHAMAS is a fully hydrodynamical simulation, with a baryonic feedback prescription based on the framework developed in the Overwhelmingly Large Simulations project \citep{eagle},  that includes realistic feedback from Active Galactic Nuclei (AGN) and supernova, radiative cooling, star formation, chemo-dynamics and stellar evolution. \cite{BAHAMAS} showed that the simulations can accurately reproduce the local stellar mass function and the gas mass - halo mass relation of galaxy clusters. We use fiducial resolution run of the BAHAMAS that simulated a large box of $400$Mpc$/h$ with a plummer equivalent softening length of $4$kpc/$h$ with $1024^3$ baryon and CDM particles in a WMAP9 cosmology \citep{WMAP9}. 

We use {\sc SUBFIND} to find and extract the 40 largest galaxy clusters (and the sub-haloes within these clusters) at a redshift of $z=0.375$, cutting out boxes of $2$Mpc in the $x-y$ direction, and $\delta z=10$ Mpc in projection. This depth is sufficient to encapsulate the additional mass from the environment of the clusters that will impact the shape of the cluster. Beyond this radius the impact of line-of-sight structures will be minimal (see Section \ref{sec:results}). The The X-ray luminosity maps are derived according to section 5.1 of \cite{BAHAMAS}. From these we measure the X-ray concentration ratio, $\Gamma=S(<{\rm100kpc})/S(<{\rm 400kpc})$ and cut at $\Gamma>0.2$ to ensure we have only relaxed clusters \citep{dynamical_state_xray}, leaving a total of 22 clusters. To derive the lensing observables we first project these into a two dimensional projected surface density and then carry out the following procedures:
\begin{enumerate}
\item {\it Simulated strong lensing observable: }
Our strong lensing catalogues are made from sources placed onto seven discrete source planes, equally spaced in redshift from $z_{\rm s} = 1$ to $z_{\rm s} = 4$. On each source plane we place a regular grid of 128x128 sources, covering an area of 96x96 arcsec$^2$, centred behind the centre of the lensing cluster. Each source is modelled as an ellipse, with an area equal to that of a circle with radius 0.5 physical kpc, and with an axis ratio drawn randomly from a uniform distribution between 0.5 and 1; a random position angle is also drawn for each source.
For each of these sources we calculate the locations and magnifications of all images of this source as they would appear in the (observed) lens plane, following the method in \citet{2020MNRAS.494.4706R}. The deflection angles due to the lens are calculated on a regular grid in the lens plane from a pixelised map of the lensing convergence (on the same regular grid) using discrete Fourier transforms (see \cite{RobertsonBAHAMAS}). Source-by-source, we find all points in the lens-plane grid that when mapped to the relevant source plane are enclosed by the boundary of the source. These points are split into contiguous sets in the lens plane  (with steps in redshift of $\delta z=0.5$), which are the individual lensed images. The magnification of an image is given by the number of lens-plane grid-points that map into the source, divided by the expected number that would map into the source in the absence of lensing. The position of the image is given by the location of the lens-plane grid-point that maps closest to the centre of the source in the source plane. The grid-spacing in the lens plane is 0.02 arc-seconds.

With a regular grid of lensed sources at different x-y positions and redshifts we randomly select a source to enter the final image catalogue. To do this we first create a magnification-biased luminosity function for each source position based on \cite{galaxyLumFunc}. We then generate a random luminosity and calculate the observed luminosity of all its respective images.  Assuming a limiting magnitude of $m < 30$, we then determine which images would be observed. By adding the condition that we must observe counter images, we have our final image catalogue.

\item{\it Simulated weak lensing observable: }
\cite{KS93} showed that the normlised projected surface density (or convergence), $\kappa$ can be related to the weak lensing shear, $\gamma$ via;
\be
\tilde{\gamma} e^{2i\theta}= \frac{ l_1^2 - l_2^2 + i2l_1l_2 }{ l_1^2 + l_2^2} \tilde{\kappa},
\ee
where the tildes denote Fourier transforms, $l$ is the wavenumber, and the shear is in the form of the complex number $\gamma=\gamma_1+i\gamma_2$.  For more on the weak lensing shear and convergence please see Appendix \ref{sec:pyRRG}. We therefore take the Fourier transform of the projected surface density and convert to a vector field. Then, by assuming a background galaxy distribution that is uniform across the sky, and a $z_{\rm source}=1.0$ and a density of $n_{\rm gal}/$arc-minute$^2=100$, we interpolate the shear field to individual galaxy positions and derive shear catalogues for each galaxy cluster. Since we want to directly compare the estimated shape from this probe (and not the expected error bars), we assume a  large background density and that each source galaxy is intrinsically circular (i.e. no intrinsic ellipticity). 
\end{enumerate}
\fig
\includegraphics[width=0.49\textwidth]{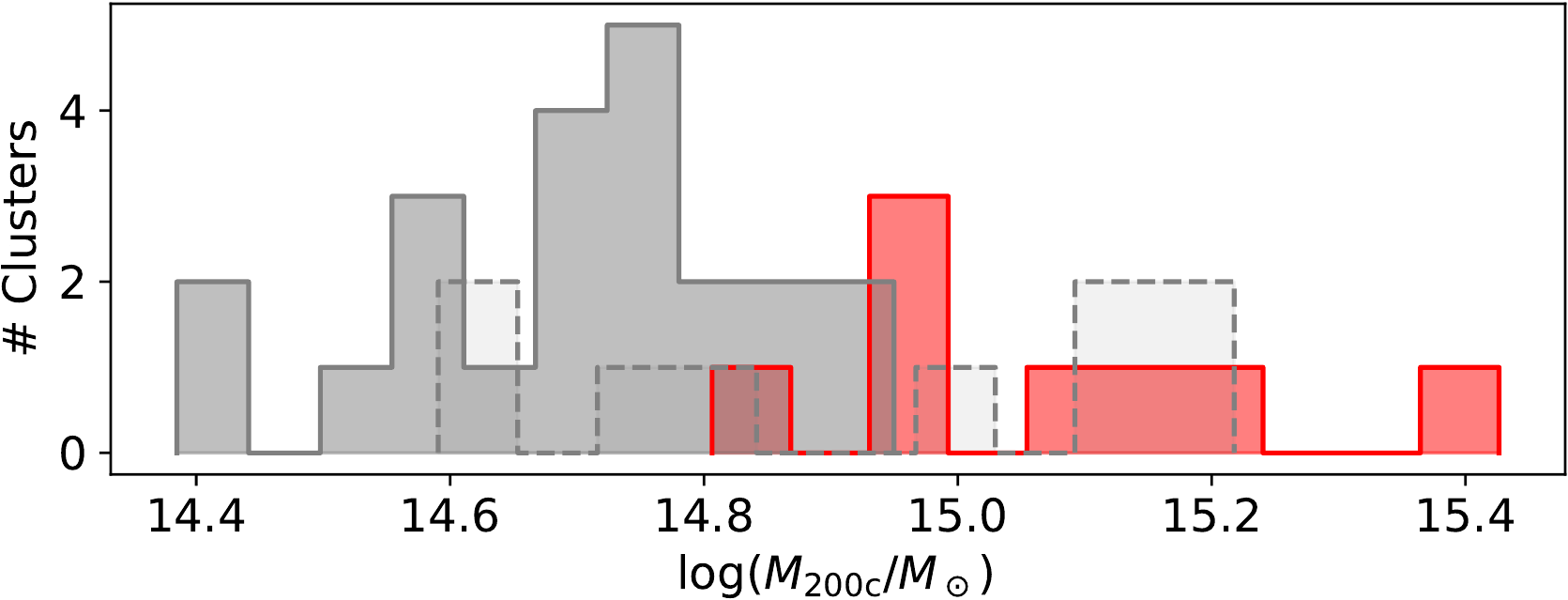} 
\caption{\label{fig:clusterMasses} The distribution of $M_{\rm 200c}$ of the observed clusters as estimated by strong gravitational lensing (red histogram) and the most massive relaxed (grey solid) and unrelaxed (grey dashed) simulated clusters in the BAHAMAS simulations at a redshift of z=0.375. }
\efig
\subsection{Observational data of massive clusters}
We use data from the Advanced Camera for Surveys (ACS) on board the Hubble Space Telescope (HST) for both the strong and weak lensing analysis, the shape of the distribution of cluster members and the shape of the BCG, and {\it Chandra X-Ray Observatory data} ({\it CXO}) for the X-ray analysis. We use the sample of galaxy clusters from \cite{Harvey_BCG} consisting of 10 strong lensing clusters from the  Local Cluster Substructure Survey \citep[][LoCuSS]{Locuss_Richard} and the Cluster Lensing And Supernova survey with Hubble (CLASH) \citep{CLASH} that have at least $10$ multiple images, ensuring sufficient constraints to measure the lensing parameters that govern the inner region of the cluster, and sufficient imaging by ACS to make a robust weak lensing measurement (i.e. one orbit in either F814W or F850LP). Furthermore, these ten galaxy clusters are required to be relaxed, with no signature of dynamical disturbance as this could bias the shape of the clusters when trying to isolate the impact of physics in the core. We quantify this by measuring their X-ray isophotal concentration, $S=\Gamma_{\rm 100kpc} / \Gamma_{\rm 400kpc}$, where $\Gamma$ is the integrated X-ray flux within a given radius. We apply a strict criteria that the cluster must have $S>0.2$ \citep{dynamical_state_xray}. Finally of the ten clusters, Abell1413 does not have sufficient optical imaging in the Advanced Camera for Surveys on {\it HST} for the weak lensing and AS1063 catalogue is currently in process from the BUFFALO survey \citep[GO-15117,][]{BUFFALO} and therefore will not be ready for another year. As such we have a final sample of eight galaxy clusters. A summary of these clusters can be found in Table \ref{tab:data}.

\subsubsection{Strong Lensing Image Selection}
For the strong lensing measurement we adopt the published multiple image catalogues of {\it confirmed} images from \cite{CLASH_zitrin}, \cite{Locuss_Richard} and \cite{a1703}. The strong lensing model requires knowledge of the redshift of the source. As such, for those images that do not have spectroscopic redshifts we match the sources to publicly available photometric redshift catalogues \citep{CLASH_photoz} and add these source redshifts as parameters in the model with the one-sigma error in the photometric redshift as a Gaussian prior.

\subsubsection{Cluster Member Selections}
In order to select cluster members we also adopt the cluster member catalogues from  \cite{CLASH_zitrin}, \cite{Locuss_Richard} and \cite{a1703}, who had identified the red sequence in order to classify the cluster members. For more please see referenced papers.

\fig
\includegraphics[width=0.5\textwidth]{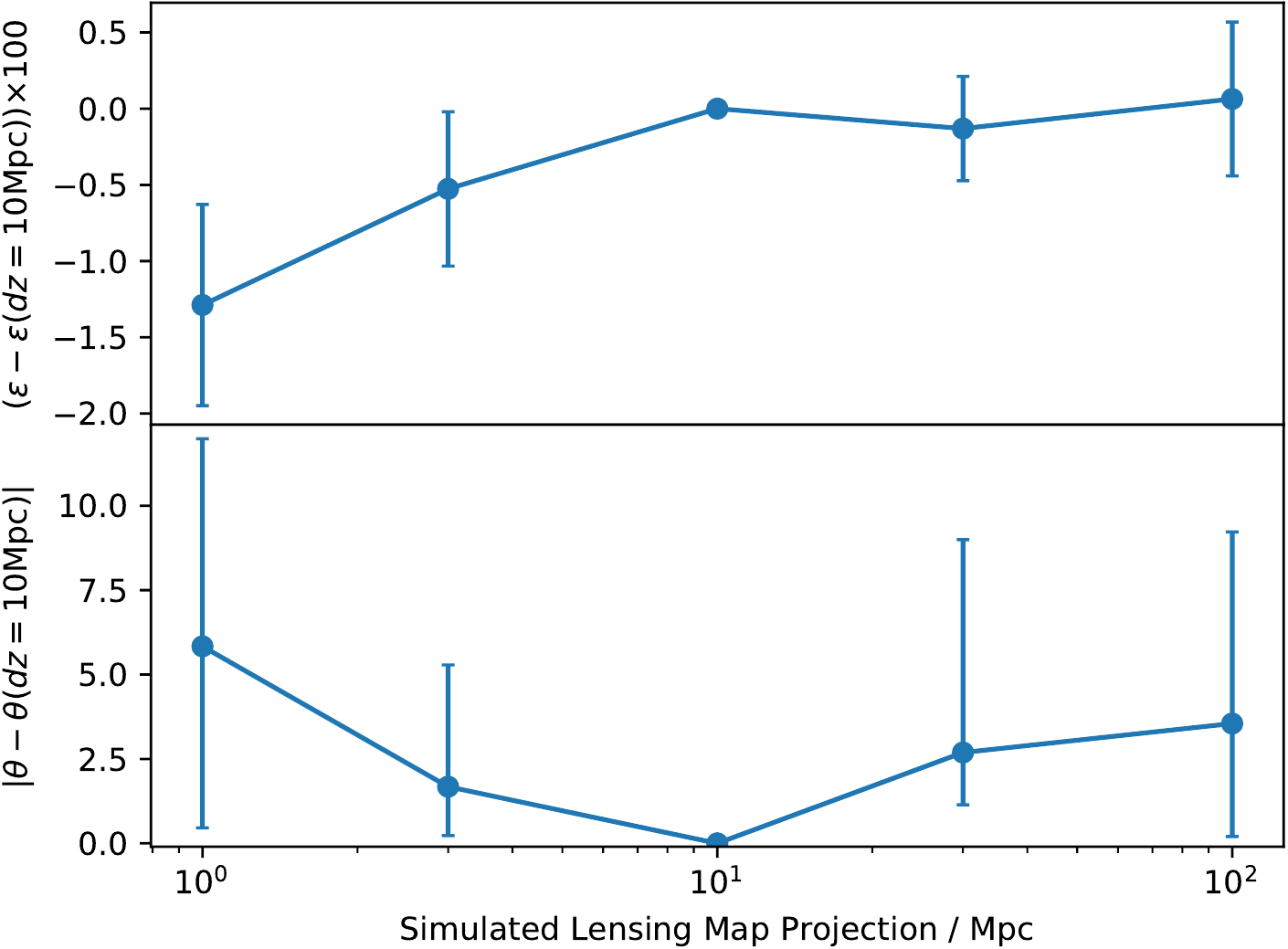}\\
\caption{\label{fig:testProjection} Impact of the cluster environment on the estimated ellipticity and position angle. We project the simulated lensing mass maps to varying depths and measure the weak lensing ellipticity (top panel) and position angle (bottom panel) relative to the fiducial projection of d$z=10$Mpc. We find that the shapes converge above at this fiducial value. We have multiplied the y-axis of the top panel by 100 for clarity.
 }
\efig

\fig
\includegraphics[width=0.5\textwidth]{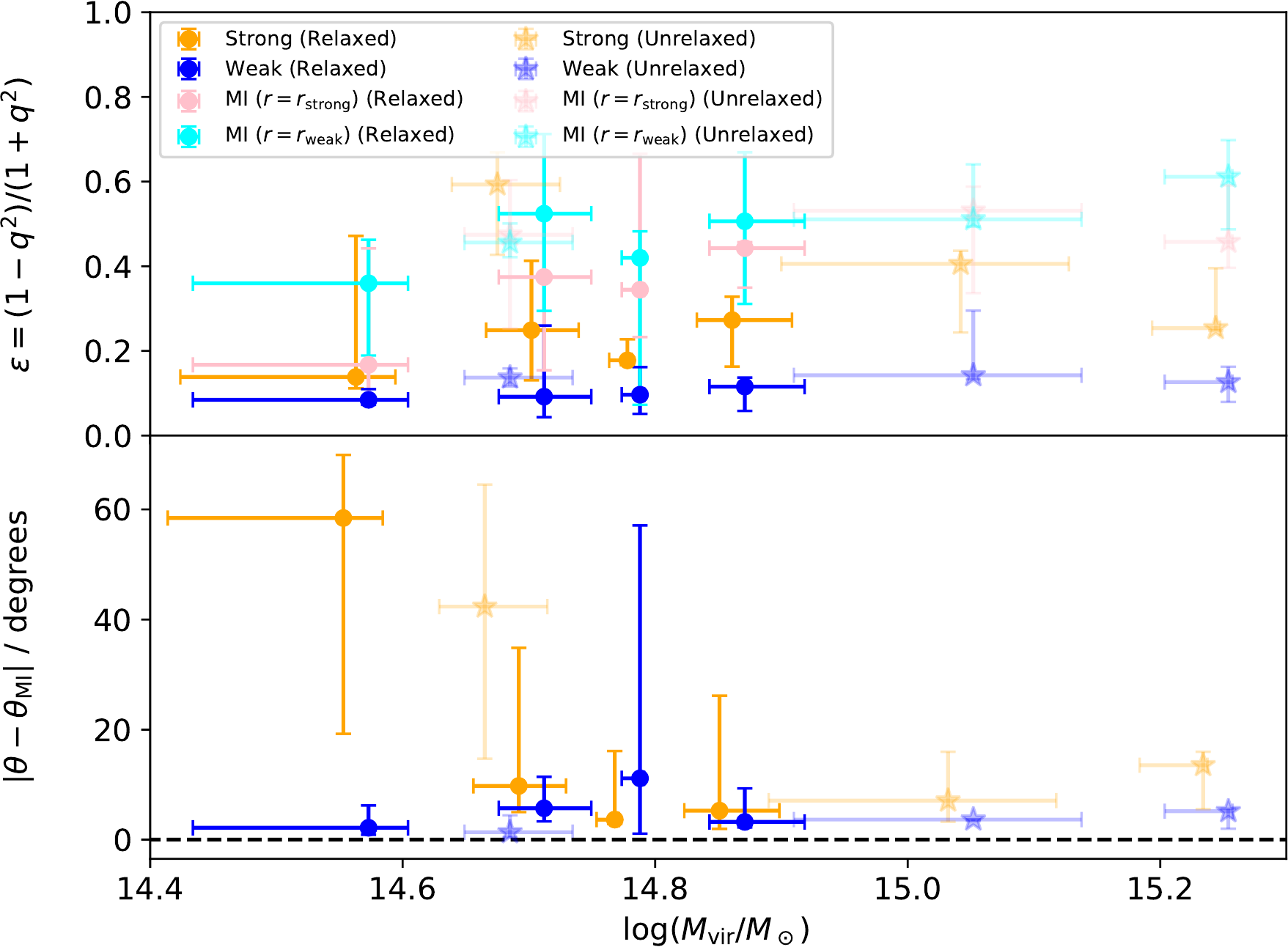}\\
\caption{\label{fig:inertiaVsObs} A direct comparison between the ellipticity (top panel) and position angle of the major axis (bottom panel) estimated from the projected moment of inertia (MI) calculated directly from the particle data at either the radius of the strong lensing (the mean radial distance of the multiple images, pink) or the weak lensing (the maximum distance of cluster members in the HST image, cyan) and the strong (orange) and weak (blue) lensing via mock observations. The solid spots represent the relaxed clusters with an xray-concentration of $\Gamma>0.2$ (see text) and the faded stars represent unrelaxed clusters with an X-ray concentration of $\Gamma<0.2$. The bottom panel shows the mis-alignment of the strong and weak lensing mock observation estimates relative to the moment of inertia at the corresponding radius.
 }
\efig

\subsubsection{Weak lensing data reduction}
We obtained raw images of each cluster with the associated reference files and re-analyse the data. We first treat each individual exposure for radiation damage due to cosmic rays inducing charge `traps' in the CCD. During read-out the trapped charge is re-released intermittently causing flux to erroneously appear along the read-out axis of the CCD. As a result we must model this Charge Transfer Inefficiency (CTI) and post-process the image in order to redistribute the charge \citep{CTI,CTI2,CTI3}. 

Following this we used the publicly available {\sc Calacs} package \footnote{\url{https://acstools.readthedocs.io/en/latest/calacs_hstcal.html}} to re-calibrate the individual raw images and then co-add them using the {\sc Astrodrizzle} package \citep{astrodrizzle}, accounting for deformations induced by the telescope. During the drizzling process we use a square kernel and a final pixel scale of 0.03"/pixel as recommended by \cite{drizzlepars}. For those exposures that are mis-aligned and taken at different epochs, we use {\sc SExtractor} to extract sources from the image and then the {\sc Tweakreg} algorithm to re-align to a common reference frame. We also use {\sc Astrodrizzle} on individual exposures to produce deformation free images required by the shape measurement process, {\sc pyRRG}. Finally, we measure the shapes of all source galaxies using the publicly available code {\sc pyRRG} (see Appendix \ref{sec:pyRRG}).

\subsubsection{CXO data reduction}
 We re-process all the available raw ACIS-I and ACIS-S Chandra data using the publicly available {\sc CIAO} tools\footnote{\url{http://cxc.harvard.edu/ciao/}} as in \cite{Harvey15} and the {\sc Chandra\_repro} script, creating new``evt2" files (the raw photon tables). This ensures that the data is cleaned with the most up-to-date reference files. Following this, we extract a region of interest and combine the exposures using the {\sc merge\_obs} script, which produces an estimate of the spatially varying Point Spread Function (PSF). We use the wavelets source finder, {\sc Wavdetect}\footnote{\url{http://cxc.harvard.edu/ciao/threads/wavdetect/}} with a filter size of $1$ and $2$ pixels to find point sources within the field that are extremely bright and not associated with the broad emission of the cluster halo. We remove the detected point sources from the field and then verify by eye, removing any residual point sources by hand. Finally we weight each pixel by the estimated exposure map constructed during the {\sc merge\_obs} script. This provides a clean, exposure time weighted flux map of the cluster.


\fig
\includegraphics[width=0.49\textwidth]{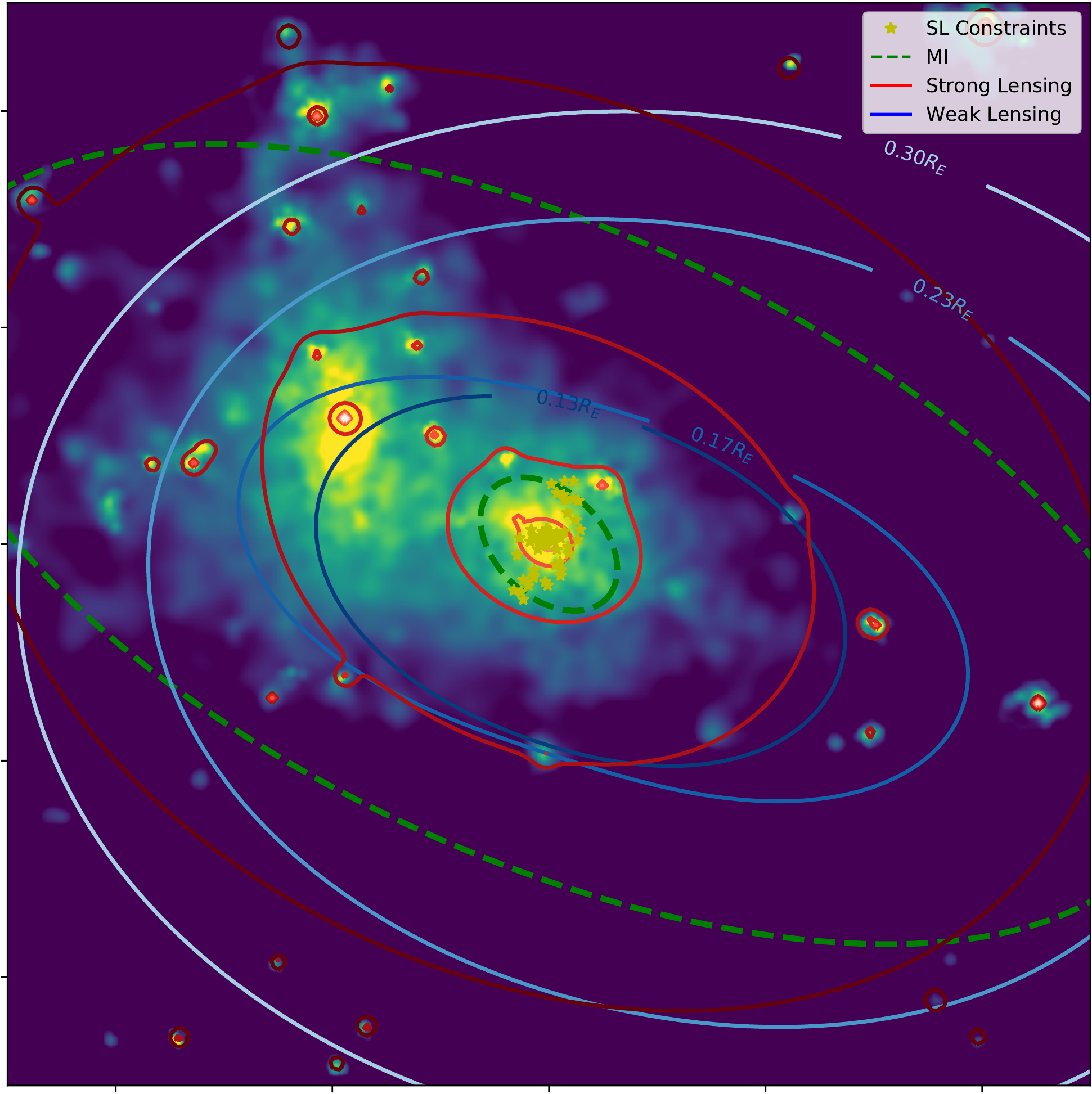}
\caption{\label{fig:failedLensing} An example when the alignment of {\it the cluster scale halo} predicted from the strong lensing model (red contours) is misaligned with the moment of inertia (green dashed). In this cluster the strong lensing constraints (yellow stars) are aligned along the 45 degree angle and will have some perturbation from the mass in the North-West of the strong lensing region, and whilst the model includes the subhalo to the East (through a scaling relation), the constraints are insensitive to this clump. As a result the strong lensing predicts a dark matter cluster halo angle of 18 degrees, whereas the moment of inertia (from the total matter) is closer to 100 degrees resulting in an $\sim 80$ degree mis-alignment. However the weak lensing model (blue) is aligned well with the broad mass distribution, agreeing with the moment of inertia.}
\efig

\subsection{Comparability of observations with simulations}
The selection function of the CLASH and LoCUSS clusters, although X-ray selected, is complicated. We select a sub-sample of these clusters that are both relaxed and have a large number of multiple images, as such their typicality can be questioned. Although we try to simplify the selection function by only choosing those clusters that have concentrated X-ray isophotes (i.e. relaxed), the fact that they have a larger number of multiple images may mean that they are more concentrated than the average cluster and more massive. Moreover, as Figure \ref{fig:clusterMasses} shows, the most relaxed massive clusters (solid grey histogram)  in the BAHAMAS simulations (at $z=0.375$) are less massive than the observed sample (red histogram), that may result in biases in the results (where we show the total mass enclosed within a radius at which the density is 200 times the critical density of the universe at that redshift). This should be taken into account when considering the generality of the conclusions (we also show the distribution of unrelaxed cluster in the dashed histogram).

\begin{table*}
\centering 
\begin{tabular}{|c|c|c|c|c|c|c|c|c|c|c|c|} 
\hline 
Cluster & Survey & $z$  & CXO (ks)  & $N_{HST}$ & WL Filter & WL Exp (ks)& $N_{\rm WL}$ & N$_{\rm I}$ & RMS & $M_{200} (\times10^{14}M_\odot)$ & $c_{\rm vir}$\\
\hline
A383 & CLASH &0.19 & 53.449 & 16 & F814W & 4.243 & 1351 &  26  & 0.75 & $16.6\pm5.1$ & $4.0\pm1.4$ \\ 
A2261 & CLASH &0.22 & 37.526 & 16 &F814W &  4.099 & 1283 &  32  & 0.68 & $6.9\pm0.6$ & $9.0\pm0.5$ \\ 
A1703 & LOCUSS & 0.28 & 82.295 & 7 & F850LP &17.800 &  1738  & 42  & 1.03  & $13.5\pm0.9$ & $4.6\pm0.2$ \\ 
A1835 & LOCUSS & 0.25 & 23.0781 & 4 & F814W & 2.360 & 1110 &  18 & 1.20 & $28.7\pm1.7$ & $3.7\pm0.2$ \\ 
MACS0744  & CLASH &0.69 & 96.111 & 16 & F814W & 8.893 & 762  & 20  & 1.52 & $9.9\pm1.4$ & $4.7\pm0.8$ \\ 
MACS1206  & CLASH &0.44 & 25.289 & 16 & F814W & 4.240 & 834   & 35  & 1.62 & $15.0\pm0.2$ & $4.8\pm0.1$ \\ 
MACS1720  & CLASH &0.39 & 71.764 & 16 & F814W & 3.988 & 960  & 17  & 1.61 & $9.8\pm0.7$ & $5.2\pm0.3$ \\ 
MACS1931  & CLASH &0.35 & 116.841& 16 & F814W & 2.991 & 596 & 23  & 0.91 & $9.7\pm0.3$ & $5.0\pm0.1$ \\ \hline
\end{tabular} 
\caption{\label{tab:data} A summary of the galaxy cluster sample. Each cluster is confirmed relaxed from its X-ray observation. We show the following:
{\it Col 1.} Name;
{\it Col 2.} Survey;
{\it Col 3.} Redshift;
{\it Col 4.} Total, cleaned effective time Chandra X-ray Observatory time;
{\it Col 5.} Number of Hubble Space Telescope available filters;
{\it Col 6.} Filter used for the weak lensing;
{\it Col 7.} Exposure time for the weak lensing filter;
{\it Col 8.} Total number of weak lensing sources available;
{\it Col 9.} Number of strong lensing multiple images;
{\it Col 10.} The Root Mean Square of the Strong Lensing fit;
{\it Col 11.} The virial mass estimated from strong lensing;
{\it Col 12.} The concentration parameter, $c_{\rm vir}$ estimated from strong lensing.
 Note that for all clusters that are part of the CLASH sample, the publicly available photometric redshift catalogues are used.}
\end{table*}

\figs
\includegraphics[width=0.19\textwidth]{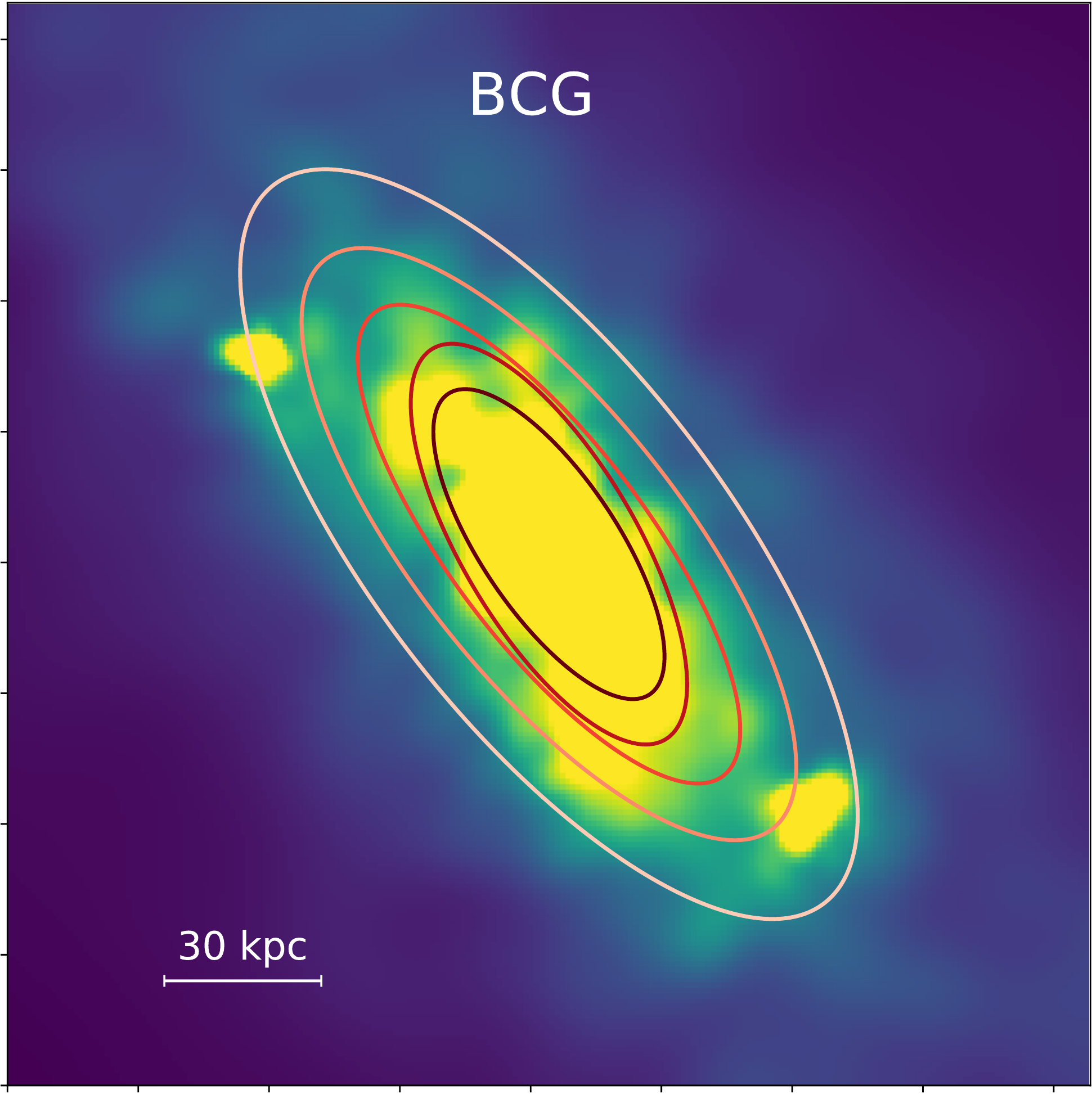}
\includegraphics[width=0.19\textwidth]{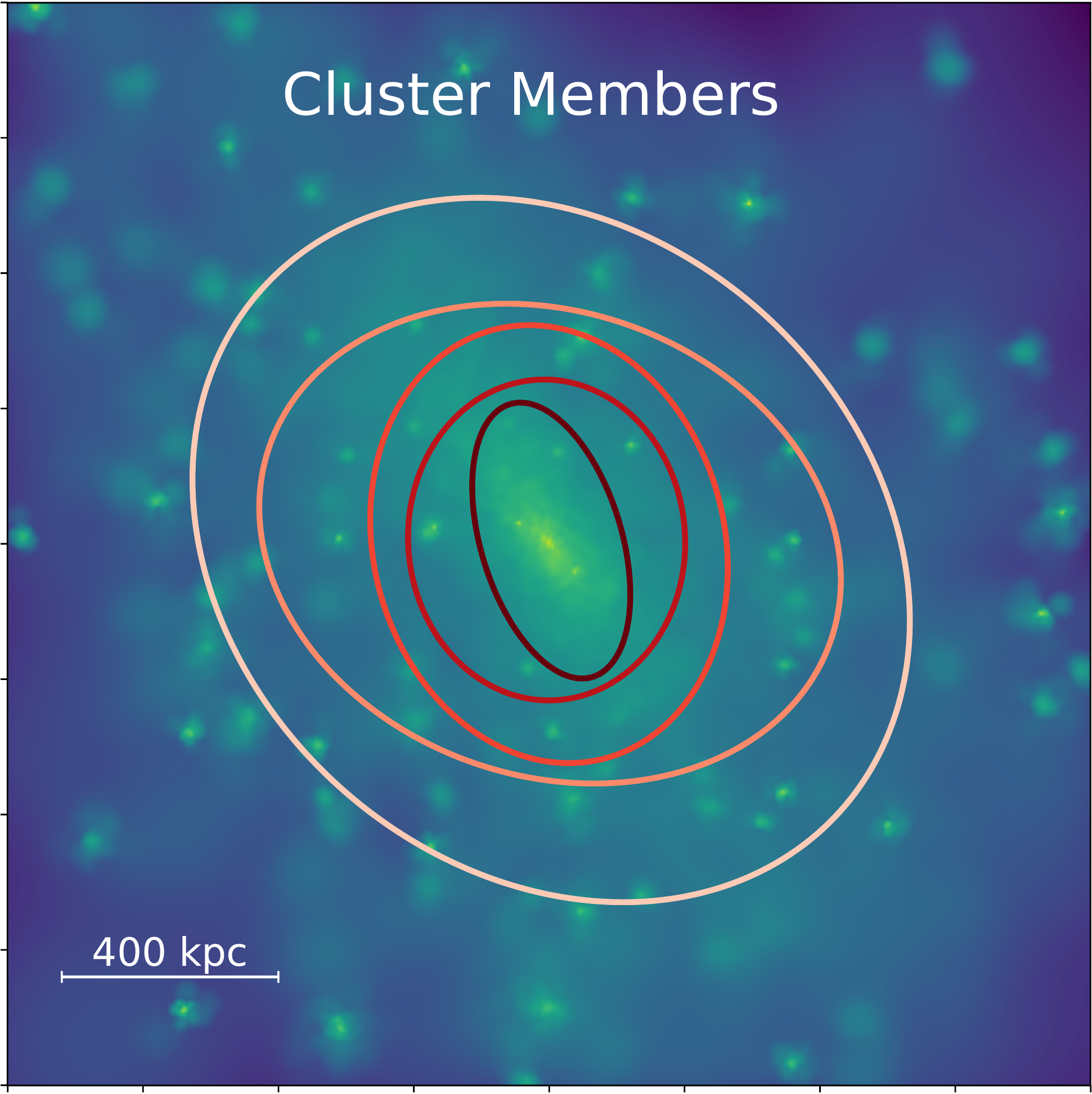}
\includegraphics[width=0.19\textwidth]{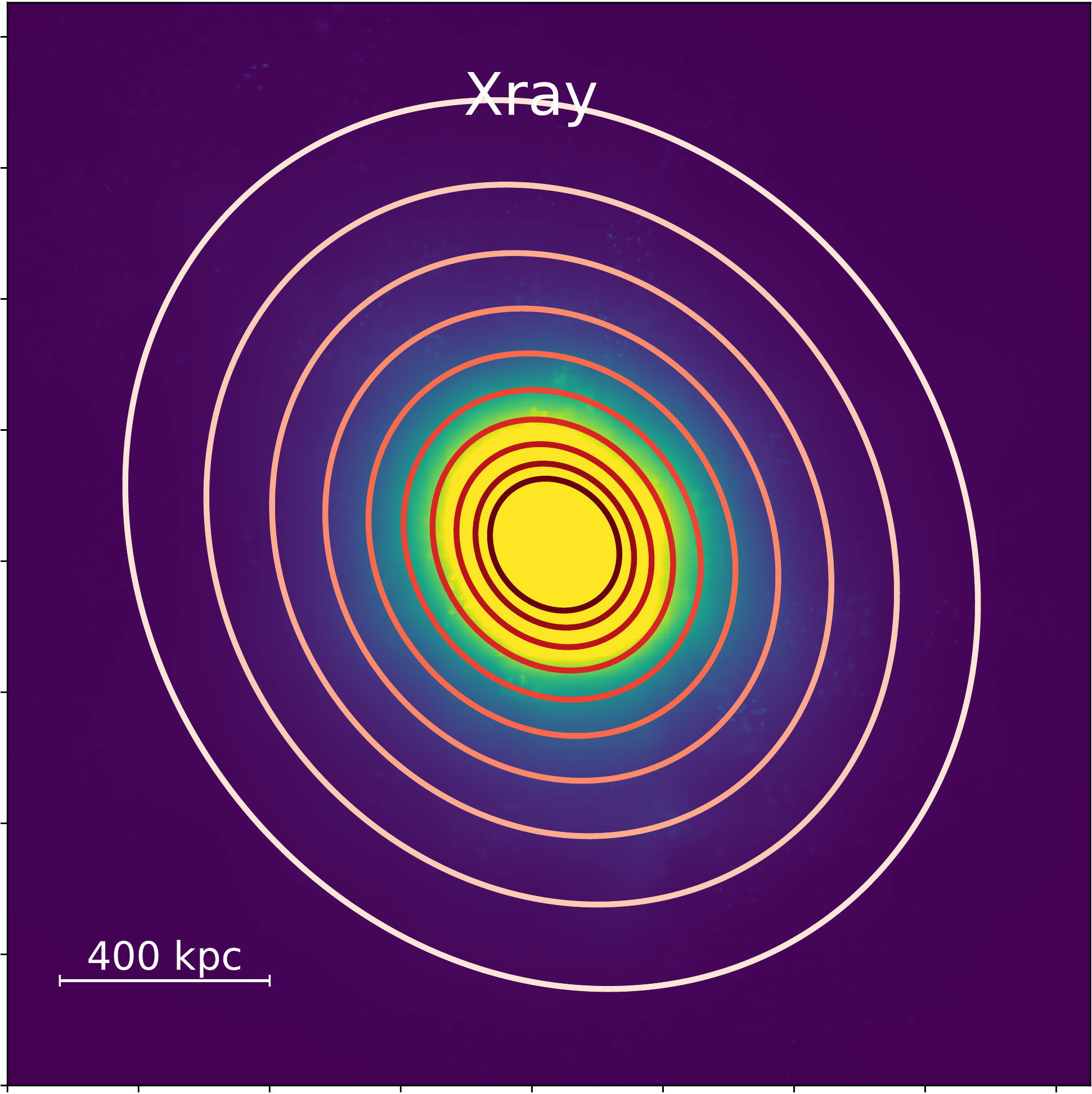}
\includegraphics[width=0.19\textwidth]{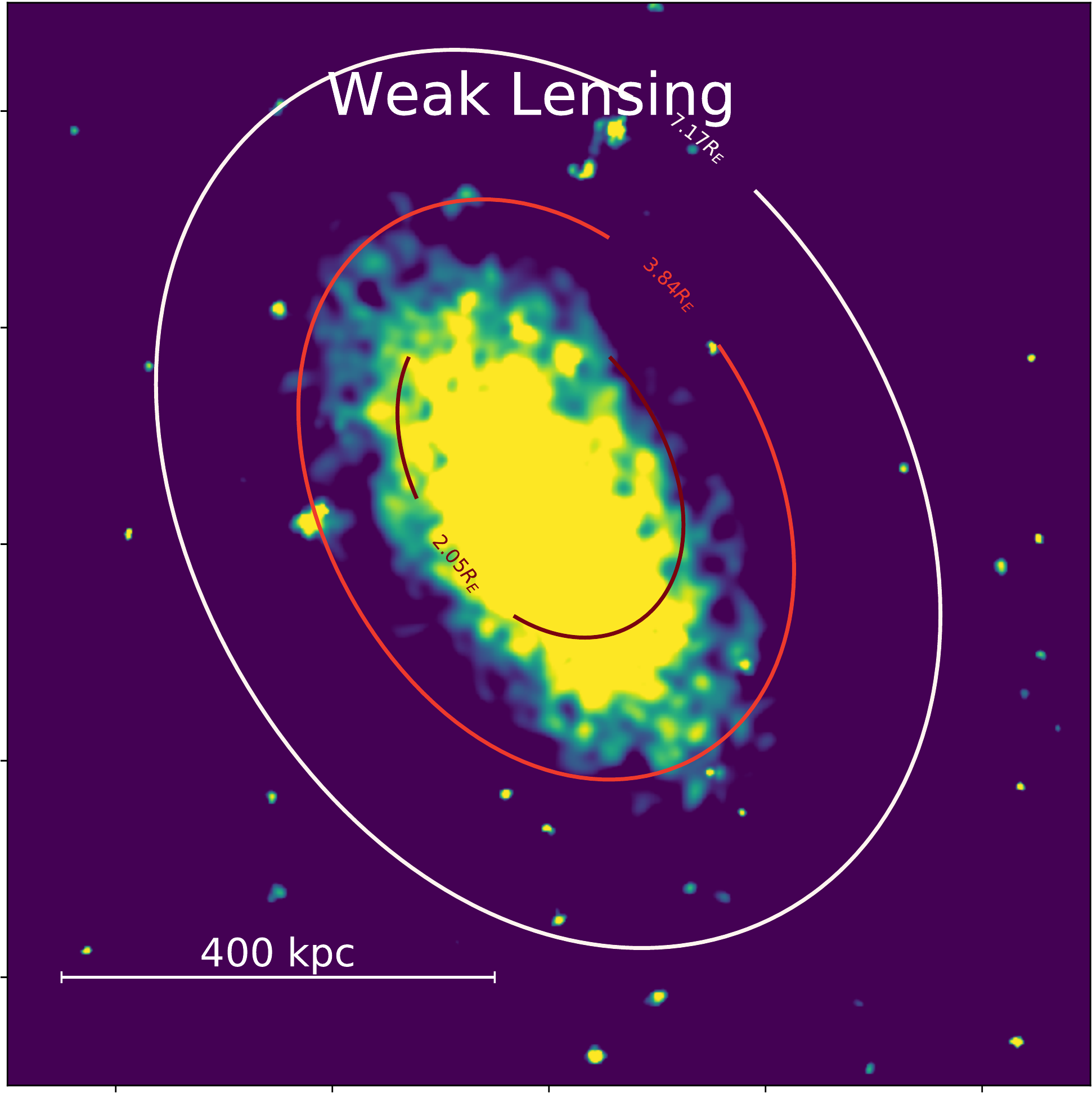}
\includegraphics[width=0.19\textwidth]{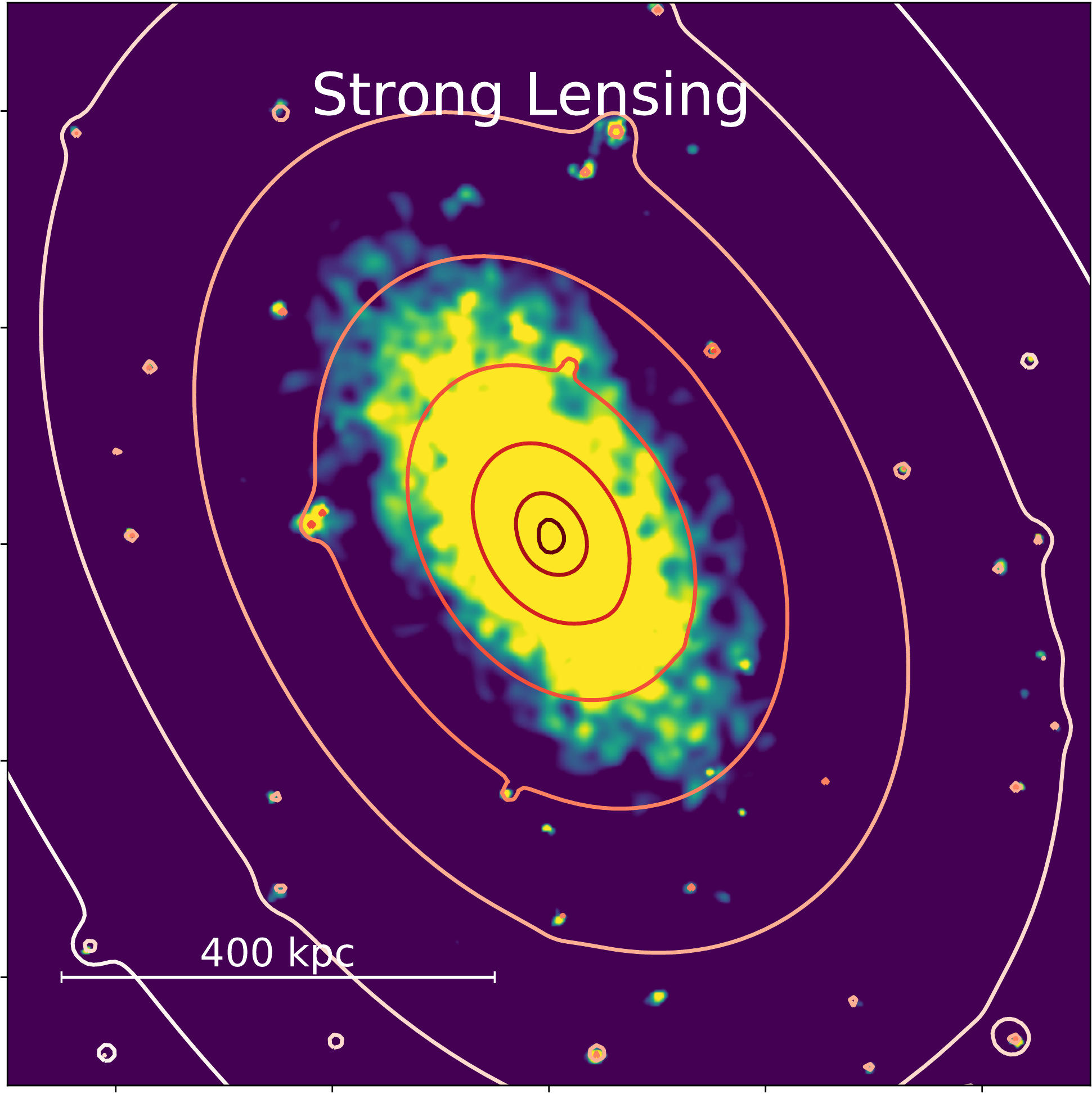} \\
\includegraphics[width=0.19\textwidth]{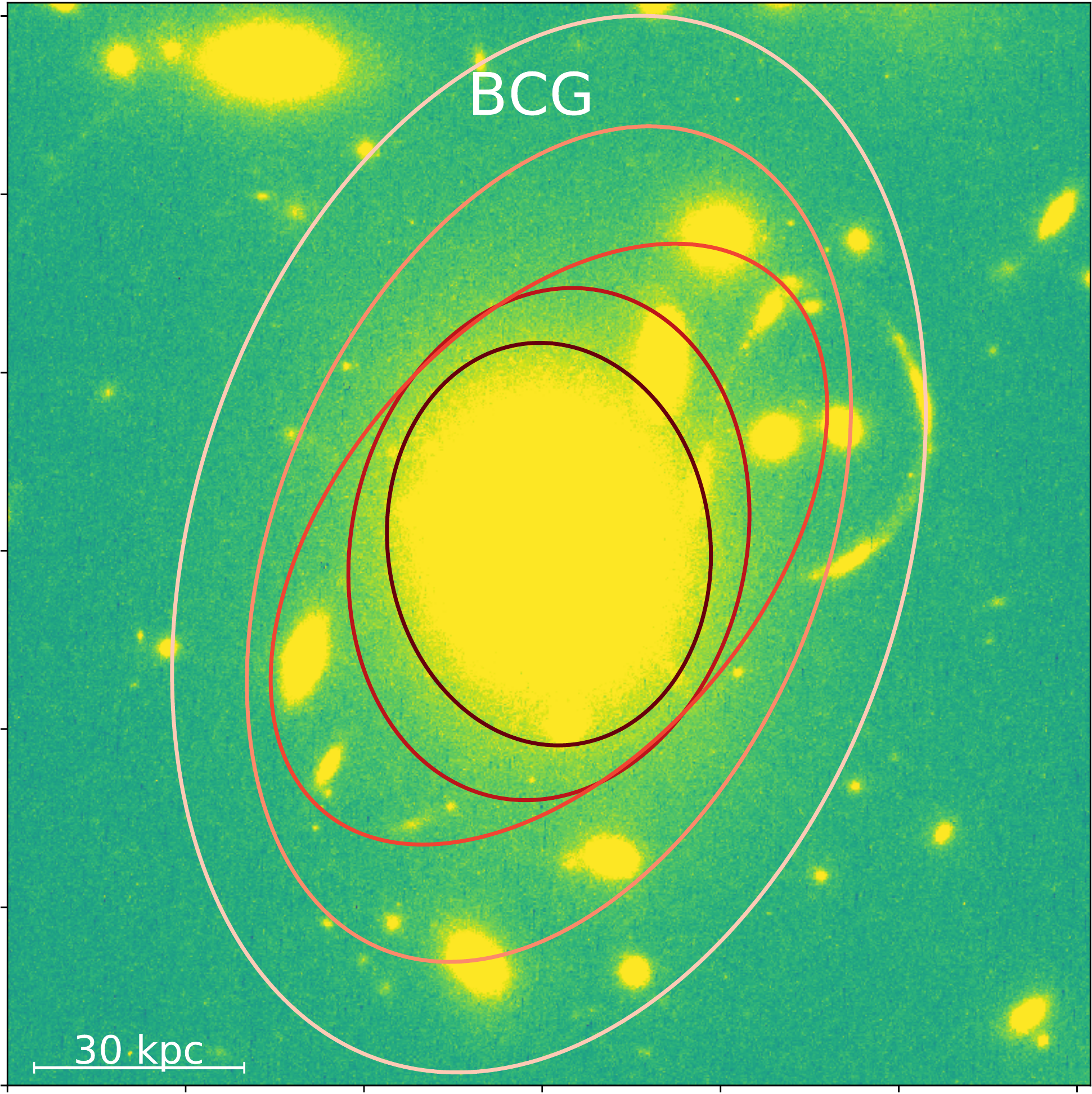}
\includegraphics[width=0.19\textwidth]{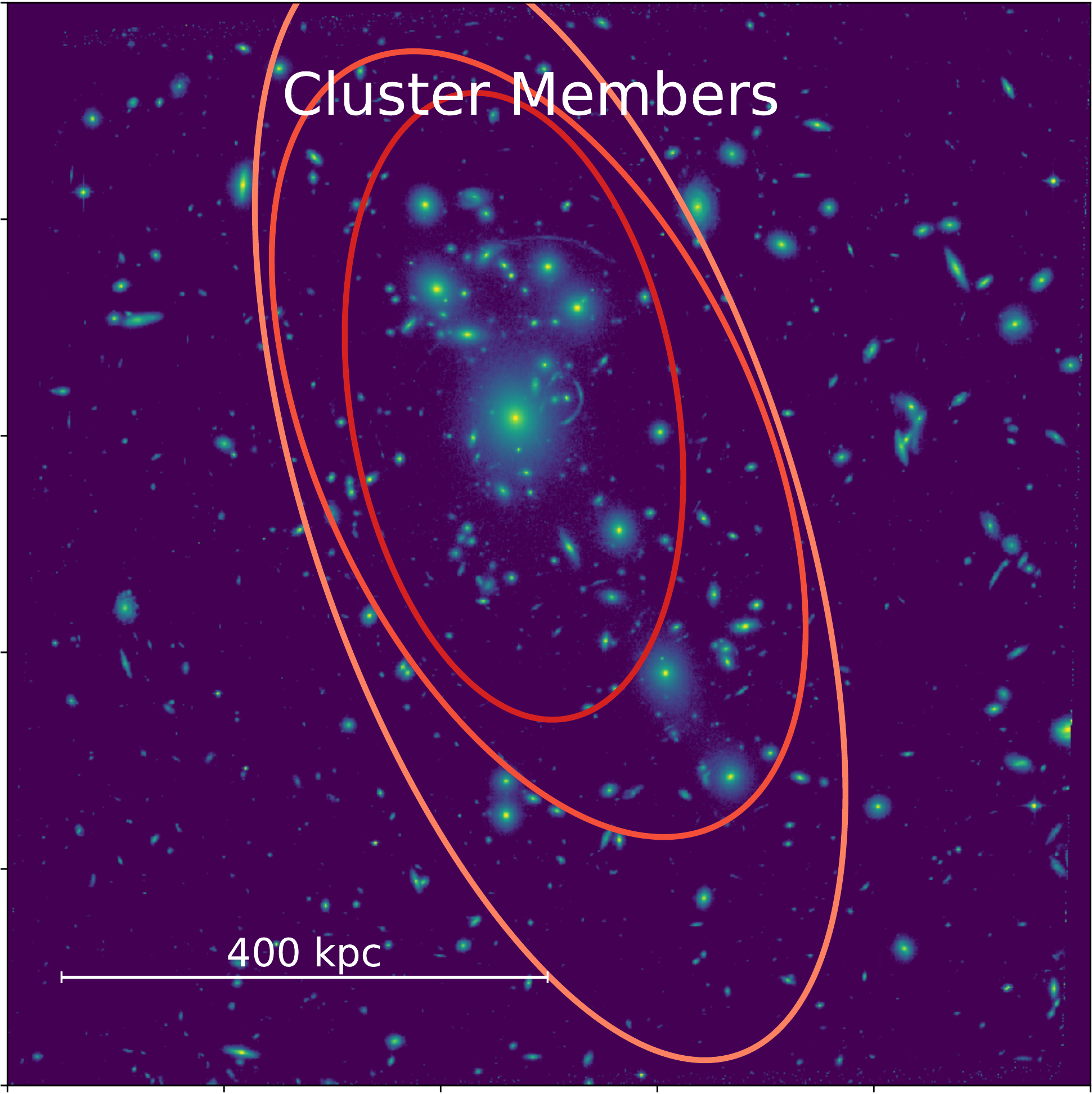}
\includegraphics[width=0.19\textwidth]{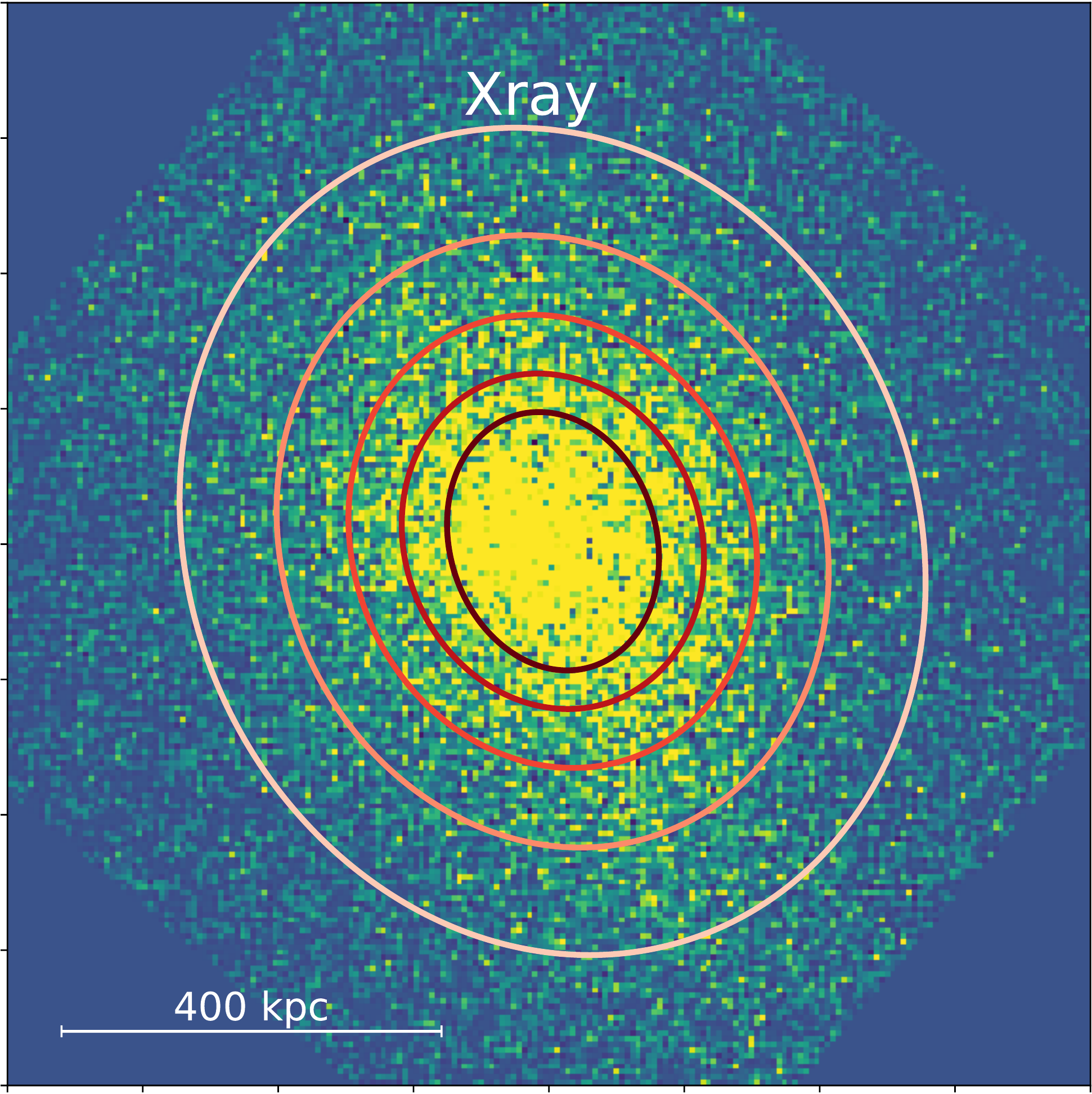}
\includegraphics[width=0.19\textwidth]{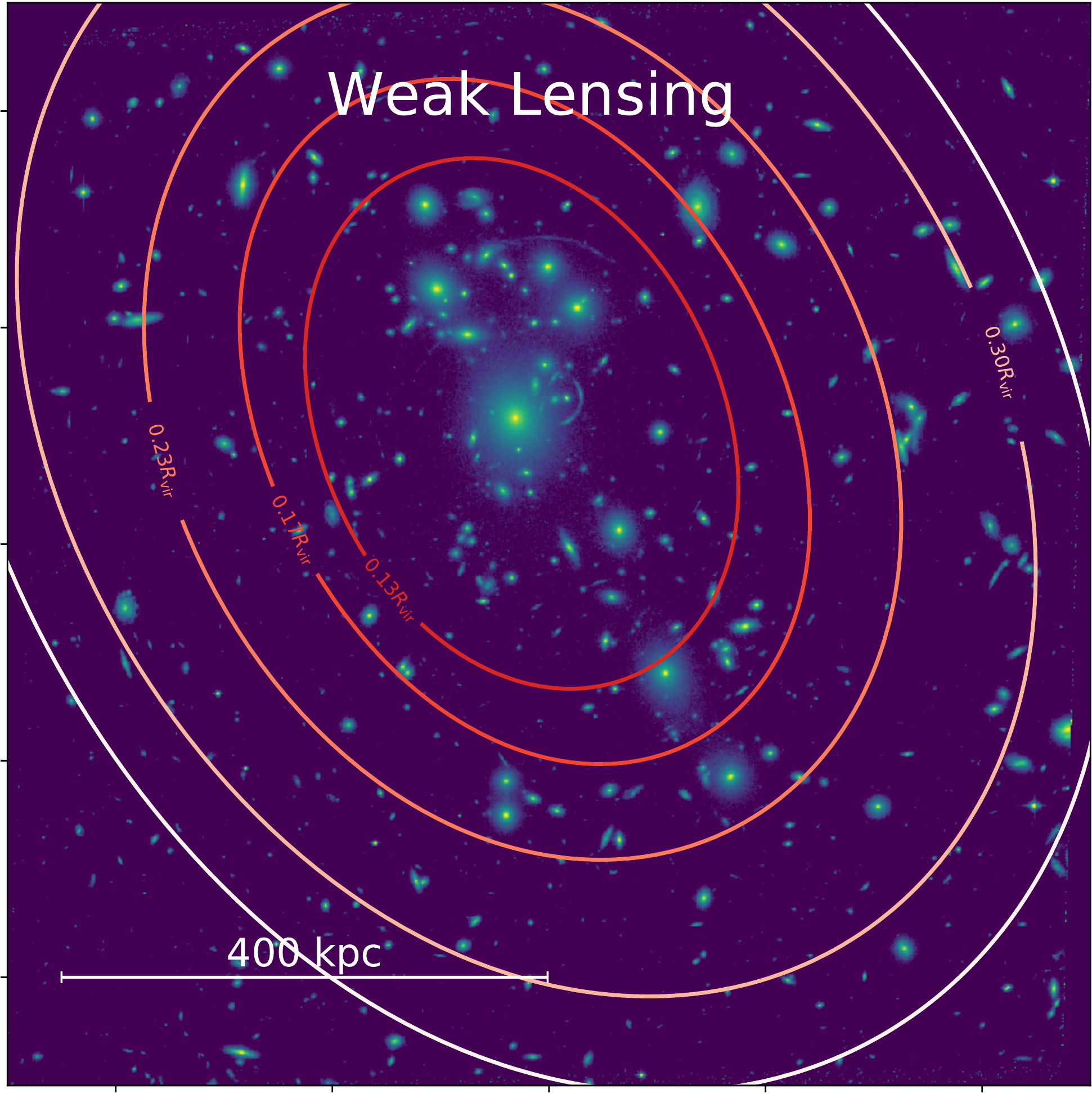}
\includegraphics[width=0.19\textwidth]{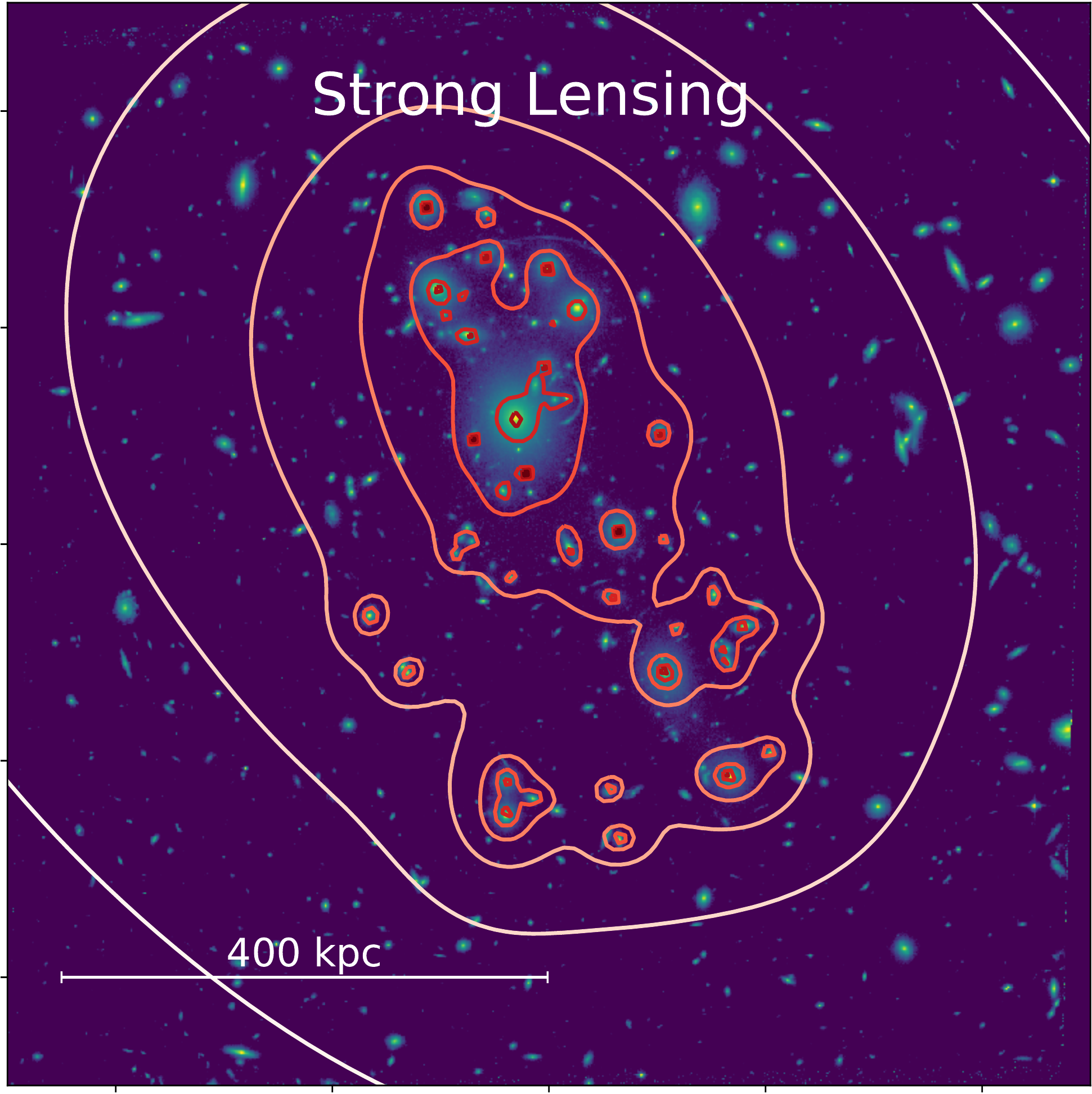}
\caption{\label{figs:dataComparison} An example of a simulated cluster from the BAHAMAS simulations (top row) and the observed cluster, A1703 (bottom row) and their best fit ellipses. In each case we show the estimated shape of the cluster for different radial cuts (decreasing shade of red). Each column shows the best fitting shapes of the Brightest Cluster Galaxy overlaid on the simulated stellar mass map (top) and HST image (bottom), the cluster members overlaid on the stellar mass map (top) and the HST image (bottom), the X-ray shape overlaid on the X-ray emission map (top) and the CXO image (bottom), the weak lensing overlaid on the total mass map (top) and the HST image (bottom) and the strong lensing overlaid on the the total mass map (top), and theHST image (bottom), respectively. Each panel has the x-y axes orientated to the  North - West and the scale is given by the white bar.
 }
\efigs

\section{Methodology}\label{sec:method}

In order to investigate our two primary questions we analyse both the twenty-two simulated clusters and the eight observed cluster with identical pipelines. Appendix \ref{sec:shapes} gives a detailed account of how we measure the shapes from each probe and how we calculate the associated errors, here we give a  brief outline. In all cases the ellipticity is defined as,
\be
\epsilon = \frac{a^2-b^2}{a^2+b^2} = \frac{1 - q^2}{1+q^2},~~\epsilon_1 = \epsilon\cos(2\theta) ~~~\& ~~~ \epsilon_2 = \epsilon\sin(2\theta),
\ee
where $a$ and $b$ are the semi-major and semi-minor axes of the ellipse respectively, $q=b/a$ is the axis ratio, $\theta$ is the angle of the major axis of the ellipse Northwards from West.
\begin{itemize}
\item{ \it Brightest Cluster Galaxy (BCG) (Appendix \ref{sec:BCG})}
We iteratively measure the flux weighted moment of inertia within a given cut radius. At each iteration we calculate the ellipticity and position angle from all pixels within an ellipse defined by the previous iteration (initialising at an ellipticity of zero). We set the minimum radial cut to be $r_{\rm cut, min}=20$kpc to avoid reaching the plummer softening length of the simulation of $4h^{-1}$kpc. Errors are calculated analytically from the fourth order moments.
\item{ \it Cluster Member Galaxies (Appendix \ref{sec:clusterMembers})} 
Derived from the mass weighted moments of the cluster member galaxies, we measure the shape using all cluster members with total stellar mass $\log(M_\star/M_\odot)>10^{10}$, matching the mass resolution of the simulations. Since the cluster members are selected by their red-sequence we also select cluster members from the simulations that have no on-going star formation. Error bars are derived via Monte Carlo realisations of the estimated shape through randomly distributing galaxy cluster members with a Poisson distribution. We measure the shape as a function of $r_{\rm cut}$, a cluster-centric radius  within which we include galaxies, and discard any bin with less than 10 cluster members.
\item{ \it X-ray Isophote (Appendix \ref{sec:xray})} 
Similar to the BCG, we derive the shape of the X-ray isophote iteratively from the flux weighted image moments, measuring the ellipticity and position angle as function from all X-ray photons within some radius cut. Error bars are derived from 100 Monte Carlo realisations of the Poisson distributed data.
\item{ \it Weak Gravitational Lensing  (Appendix \ref{sec:weak})}
We use the  shape measurement code {\sc pyRRG} and mass mapping algorithm, {\sc Lenstool} to estimate the ellipticity of the cluster from weak gravitational lensing. {\sc Lenstool} is a parametric fitting procedure, whereby we fit Navarro, Frenk and White \citep{NFW} density profiles to the data. The error bars are derived from the width of the posterior during the MCMC fitting procedure.
\item{ \it Strong gravitational lensing (Appendix \ref{sec:strong})}
The shape of the mass distribution as estimated from strong gravitational lensing using the mass mapping tool {\sc Lenstool}. We fit an NFW profile for the cluster scale halo along with Pseudo Isothermal Elliptical Mass Distriubtions (PIEMD) to each cluster member. We also assume that the cluster members lie of the fundamental plane with a constant mass-to-light ratio. Error bars are derived through 20 Monte Carlo realisations of the estimated shape and the multiple image positions (since the width of the posterior was found to be a biased estimator of the error bar in \cite{Harvey_BCG}).
\end{itemize}

\section{Results}\label{sec:results}

Following the measurement of the ellipticity of each cluster in the observed and simulated sample we present our results.  Before we address our first question we first test the environmental impact of substructures on the weak lensing shape estimate of the clusters. To do this we project the extracted lensing maps from the simulation box to varying depths, d$z$. We show in Figure \ref{fig:testProjection} the estimated ellipticity relative to the fiducial depth of d$z=10$Mpc in the top panel and the misalignment angle of the cluster relative to the estimate at the fiducial projected depth of  d$z=10$Mpc in the bottom panel. We find that at small projected depths the ellipticity is under-estimated and that the ellipticity estimates converge at the fiducial value of  d$z=10$Mpc and thus this encapsulates the total information in the environment, and that the misalignment angles remain consistent within the uncertainty.

Having justified the projection depth of our simulations, we begin by addressing our first question: {\it ``Is the ellipticity calculated from the projected moment of inertia derived directly from the particle data in simulations a good estimator of the shape derived from strong or weak lensing''}. Figure \ref{fig:inertiaVsObs} shows the mass-binned results from all 40 clusters in the simulated sample, each point showing the median, 16\% and 84\% of the distribution, the solid spots represent all those relaxed clusters that have an X-ray concentration $\Gamma > 0.2$ and the faded stars all unrelaxed clusters with $\Gamma<0.2$.  For more on how we calculate the moment of inertia please see Appendix \ref{sec:MI}. The top panel shows the ellipticity from four estimators as a function of mass, they include: {\it 1. Pink, } the 2D ellipticity calculated from the projected moment of inertia (MI) measured on {\it all} particles within a mean radius at which the mock strong lensing is measured, {\it 2. Cyan, } the same ellipticity calculated from the moment of inertia using all particles within the outer most radius at which the mock weak lensing is measured (i.e. $r < 0.4r_{\rm vir}$), {\it 3. Orange, } the ellipticity of the cluster scale halo estimated by mock strong lensing observations and {\it 4. Blue, } the ellipticity estimated by mock weak lensing observations. The bottom panel shows the mis-alignment of the position angle of the major axis between the two lensing estimates and the moment of inertia calculated at their respective radii (i.e $\theta_{\rm strong} - \theta_{\rm MI, strong}$ and  $\theta_{\rm weak} - \theta_{\rm MI, weak}$ ). 

In general we find that the ellipticity estimates from the strong and weak lensing differ from the estimate derived directly from the particle data, whereby the  strong lensing and projected inertia tensor differ by a factor of $\langle\epsilon_{\rm S}/\epsilon_{\rm S, MI}\rangle= 0.64_{-0.04}^{+0.05}$ and the weak lensing under-estimates the ellipticity by a factor of $\langle\epsilon_{\rm W}/\epsilon_{\rm W, MI}\rangle=0.23_{-0.01}^{+0.01}$. However, the strong-lensing estimate for the disturbed clusters seem to be more robust, with a smaller bias of $\langle\epsilon_{\rm S}/\epsilon_{\rm S, MI}\rangle=0.8_{-0.1}^{+0.2}$ for the entire unrelaxed sample. This is counter-intuitive since it would be easier to model a relaxed cluster yet they remain biased. 

 We find that the position angles in the bottom panel of the respective observable are well-aligned, with a mean mis-alignment angle (in the relaxed sample) of $\langle |\theta_{\rm S} - \theta_{\rm S, MI}|\rangle=8_{-2}^{+13}$ degrees  and $\langle |\theta_{\rm W} - \theta_{\rm W, MI}|\rangle=4.5_{-0.9}^{+2.0}$ degrees. However, the misalignment angle of the strong lensing is dominated by the first low-mass, should we remove this we find that the strong lensing agrees much better with a mis-alignment angle of  $\langle |\theta_{\rm S} - \theta_{\rm S, MI}|\rangle=5_{-1}^{+2}$degrees. We find that for the unrelaxed sample the strong lensing has a large mis-alignment angle  $\langle |\theta_{\rm S} - \theta_{\rm S, MI}|\rangle= 13_{-3}^{+11}$ degrees whereas the weak lensing, even in the unrelaxed sample is good agreement with the moment of inertia  $\langle |\theta_{\rm W} - \theta_{\rm W, MI}|\rangle=3.7_{-0.9}^{+0.6}$ degrees. We verify by eye those clusters that have a large difference in the angle and find that massive structures in the core of the cluster can induce huge mis-alignments. For example Figure \ref{fig:failedLensing} shows an example cluster from the BAHAMAS simulations whereby the cluster scale dark matter halo as predicted by strong lensing (red contours) is misaligned with the moment of inertia (green dashed ellipse)  by $\sim 80$ degrees, whereas the weak lensing (blue) is well aligned. We find that this is due to the distribution of strong lensing constraints (yellow stars) do not fully probe the inner region, and are likely perturbed by structures within the core. We see a halo in the North-West of the inner region that is likely biasing the strong lensing model, whereas the large halo to the East is outside the constraints and therefore not sensitive to this (although this halo will be included in the model through an assumed mass to light scaling relation). As such, it is important that when comparing strong lensing estimates to the moment of inertia the entire distribution of mass must be taking into account and not just the cluster scale halo.

Finally, we estimate the cut radius for which the moment of inertia best matches the mock weak and strong lensing. To do this we find that effective radii of the moment of inertia at which the mis-alignment angle with respect to the weak and strong lensing is the smallest, i.e.
\be
r_{\rm eff} =r_{\rm cut}[\min\{ \theta - \theta_{MI}(r) \}].
\ee
We find that the radius at which the alignment with respect to the weak and strong lensing is a minimum, match the radius at which the weak and strong lensing is measured. i.e. the moment of inertia best matches the weak lensing when measured at the same radius as the weak lensing (and same with the strong lensing).

Clearly in the real, observed case, the unrelaxed clusters would be modelled by multi-halo components, with each cluster carefully studied. In this case we have simply modelled each one with a a single cluster halo. However, these findings do still reinforce the need to carry out complete end-to-end comparisons between observations and data.

Following this investigation we move to directly comparing the observed sample with the simulated sample. As an example of both pipelines, we show the results from a single cluster in each sample in Figure \ref{figs:dataComparison}. The top row shows a simulated cluster and the bottom the galaxy cluster A1703. The first column shows the estimated shape of the BCG from the stellar mass map (first row) and the HST optical image (second row). The second column shows the estimated shape of the cluster member galaxies, once again from the stellar mass map (first row) and the HST optical image (second row). The third column shows the estimated shape of the X-ray isophote overlaid on the X-ray emission map from the simulation (top) and the CXO (bottom), the fourth column shows the best fit weak lensing estimate overlaid on the total mass map (top) and the HST image (bottom) and the fifth column shows the best fit strong lensing model overlaid on the total mass map (top) and the HST image (bottom). Each panel has the x-y axes orientated to the  North - West and the scale is given by the white bar. Each ellipse corresponds to a larger cut radius (except for the strong lensing which does not have any cut). We see that for these two examples the halo shapes as measured from different observational probes correlate with one another.

With these models we address our second question: {\it ``Is there any evidence for a radial dependent ellipticity in relaxed clusters?''} We start by showing Figure \ref{fig:differentComponentsRadius}, which presents the ellipticity of each observational probe as a function of the radius cut (normalised to the virial radius of each cluster). The top panel of each figure shows each individually observed cluster (with the legend above the figure), and then the median, 16\% and 84\% percentile of the simulated sample in the grey shaded regions (with the dashed line showing the median value). The bottom panel of each figure shows the distributions of ellipticity that are relative to the individual cluster at a specific (arbitrary) radii. This allows us to compare trends in the changing shape of a cluster. The top of each panel denotes the respective probe, with BCG in the top left, cluster members in the top right, X-ray in the bottom left and weak lensing in the bottom right. We do not show strong lensing since it has no measured radial dependence, plus the weak lensing and cluster member estimates for each cluster are offset from one another for clarity. We find in general that the ellipticities in the simulations match that of the observations, however we do find that the clusters MACSJ1206 and MACSJ1931 seem to be outliers with respect to both the observed and simulated samples. Specifically we find:
\begin{itemize}
\item {\it BCG, Top Left of Figure \ref{fig:differentComponentsRadius}}: The ellipticiites of the observed sample matches that of the simulated, however MACS1206 remains outside the 64\% region of the simulations. The error bars of A383 and A1835 are are a result of the large amounts of substructure within the BCG isophote. We also find that the central parts of the simulated BCGs tend to be slightly more elliptical than the outer regions. 
\item {\it Cluster Members, Top Right of Figure \ref{fig:differentComponentsRadius}}: We find that the simulated sample are on average consistent with the observed sample. Interestingly A1703, which exhibits no strong ellipticity in other probes has a  elliptical distribution of cluster members, whereas MACS1206 shows the opposite with a  circular distribution. The median of the simulated halos suggest that the centre of the cluster has a more spherical distribution of cluster members.

\figs
\includegraphics[width=\textwidth]{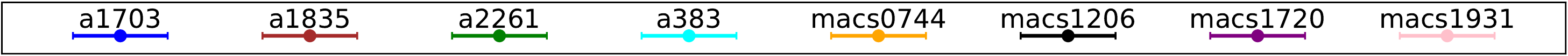}\\
\includegraphics[width=0.49\textwidth]{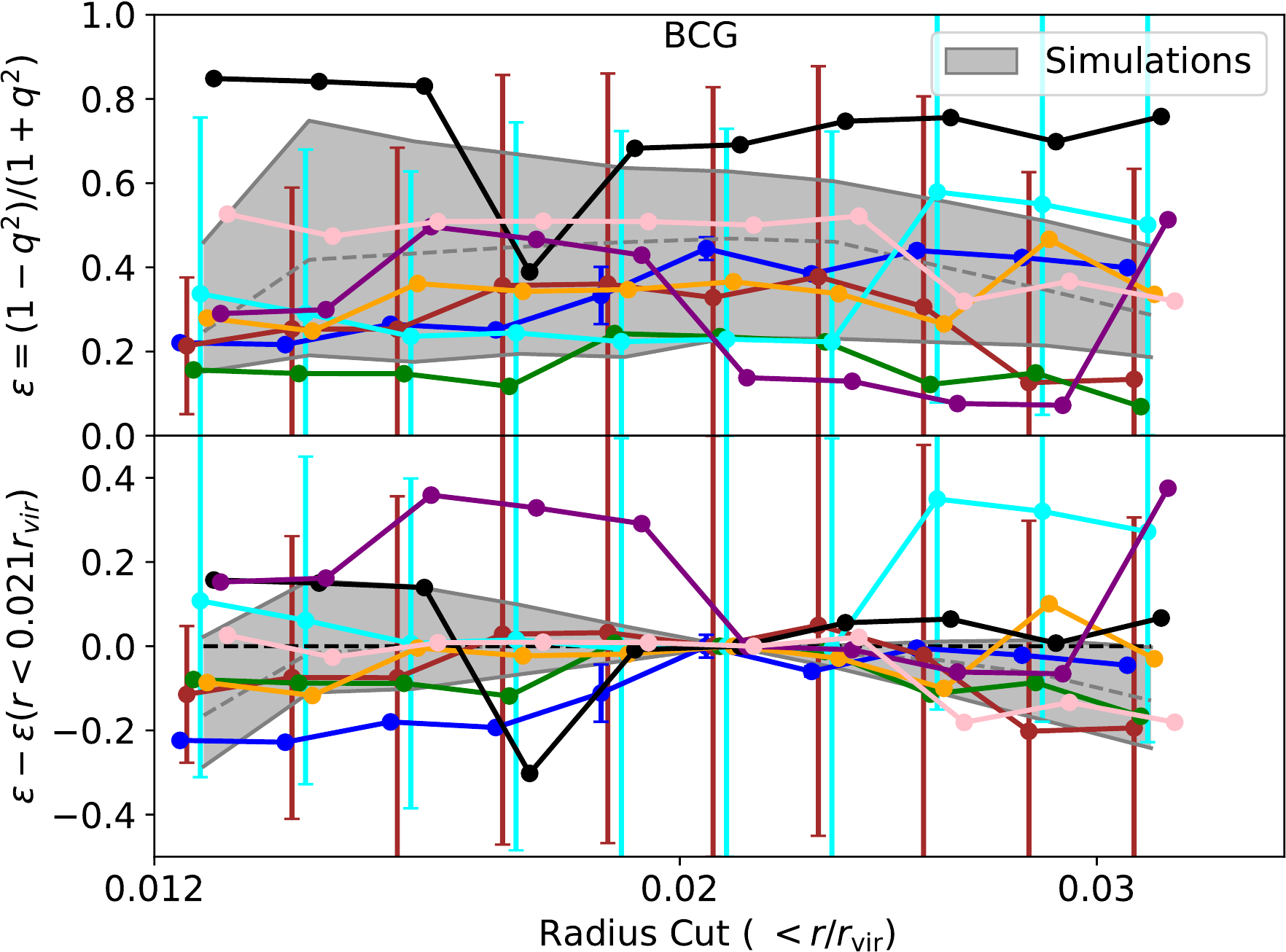}
\includegraphics[width=0.49\textwidth]{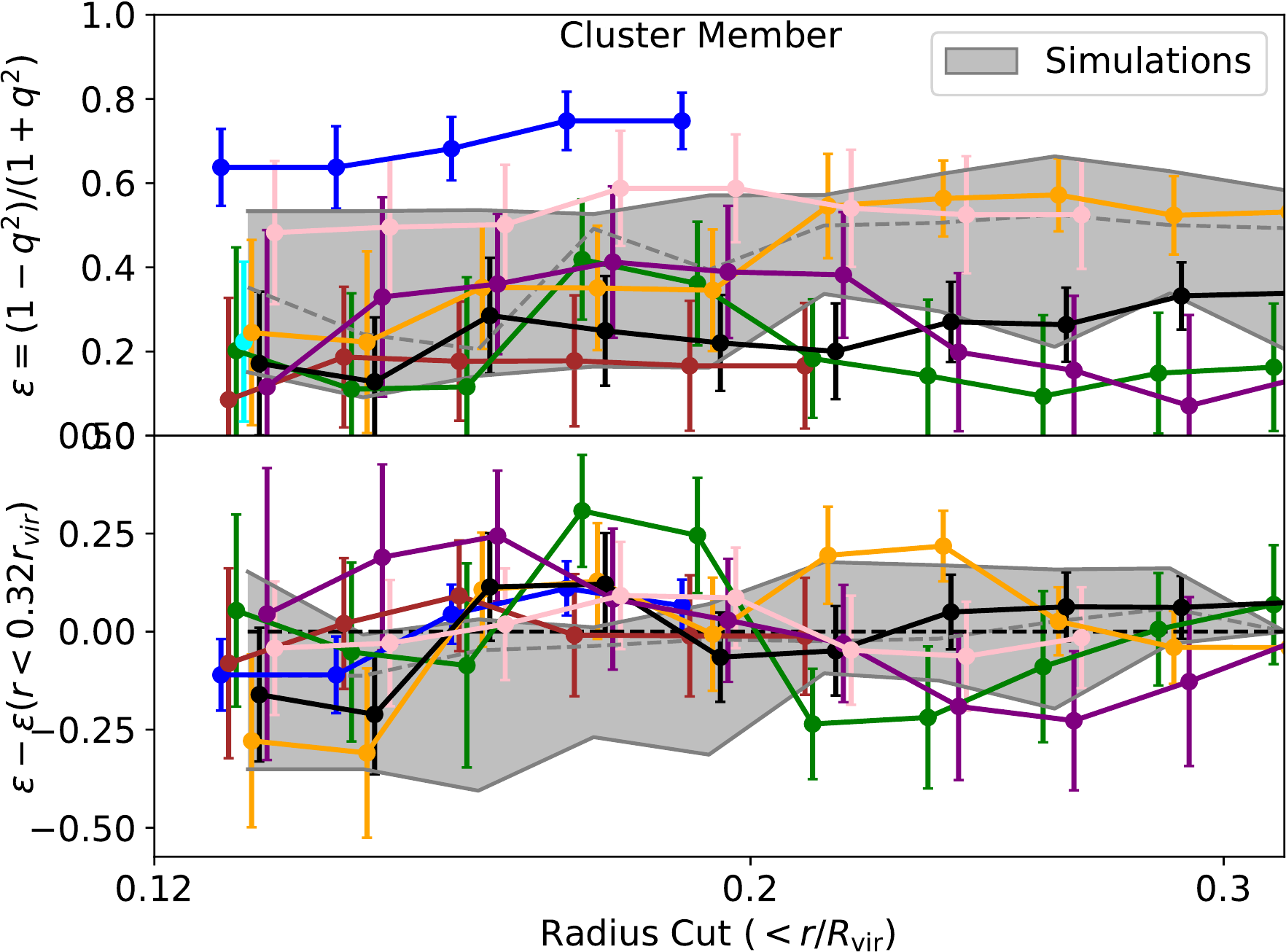} \\
\includegraphics[width=0.49\textwidth]{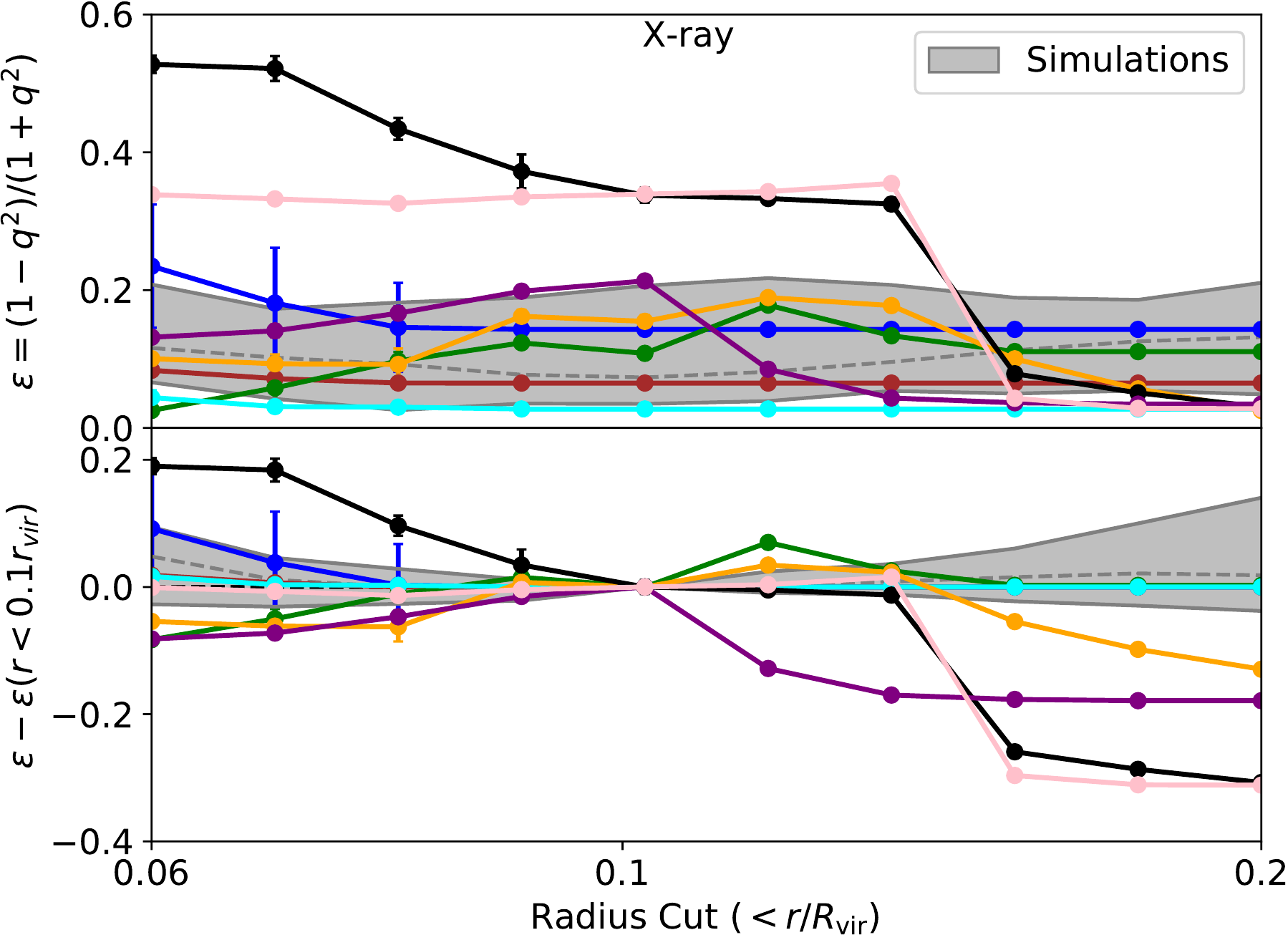}
\includegraphics[width=0.49\textwidth]{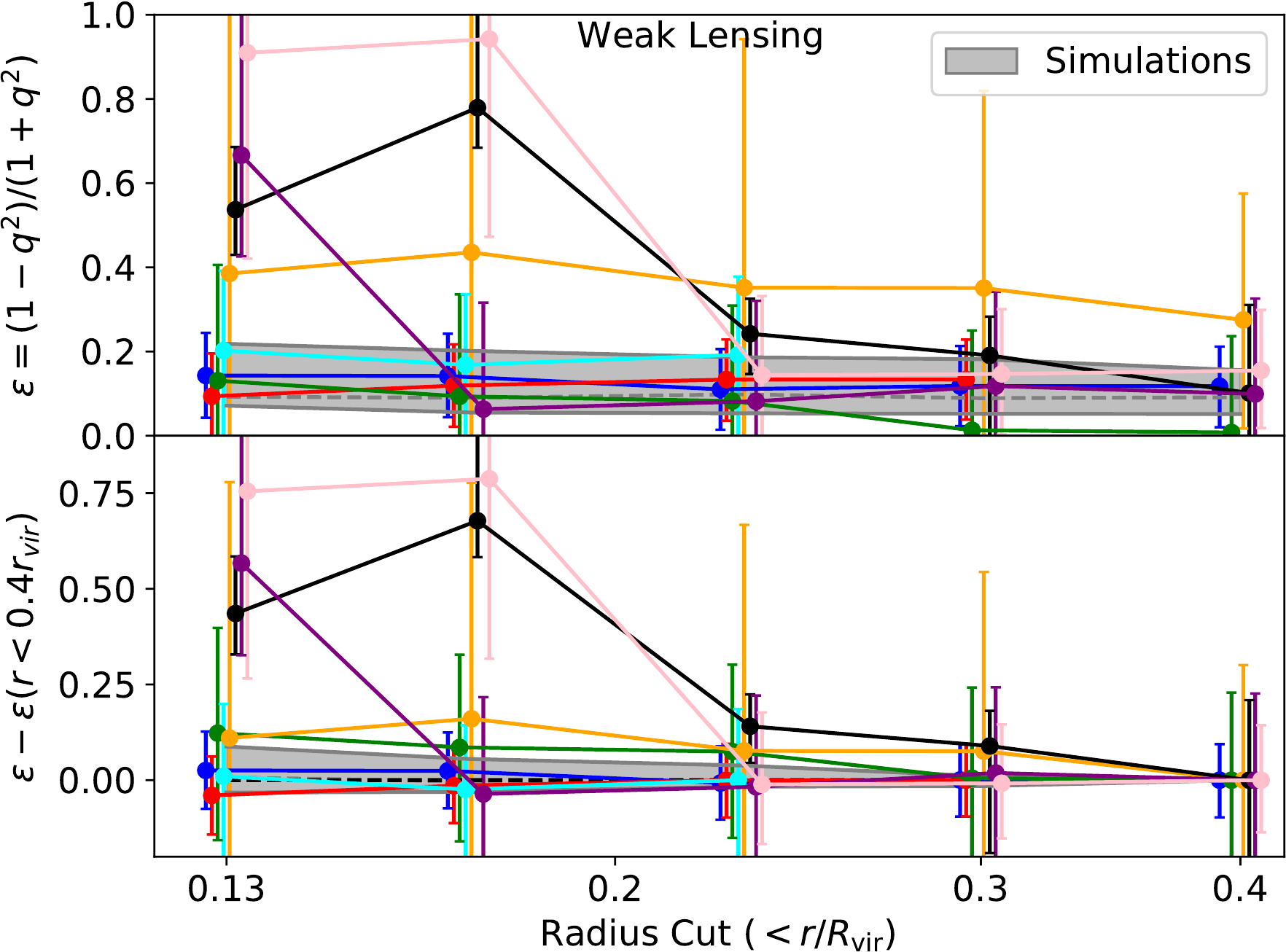} 
\caption{\label{fig:differentComponentsRadius} We show the comparison between the observed data set and the BAHAMAS simulations of the ellipticity ($\epsilon=(1-b^2/a^2)/(1+b^2/a^2)$) as a function of the radius cut. Each figure shows a different probe of the galaxy cluster (Clockwise from top left: BCG; Cluster Members; Weak Lensing, X-ray). Within each figure, the top panel shows the absolute  of individual clusters, with the legend for each cluster at the top of the main figure. The grey shaded region in each top panel shows the median and the 16\% and 84\% of the simulated. The bottom panel in each case shows the distributions of each cluster relative to some fixed radius. }
\efigs


\item {\it X-ray, Bottom Left of Figure \ref{fig:differentComponentsRadius}}: The third panel showing the X-rays shows how inner regions of the two clusters, MACSJ1931 and MACSJ1206 are both inconsistent with the simulated halos, with MACS1206 showing a clear radial dependence not seen in the simulations. We also find that the observed X-ray distribution exhibit a stronger dependency on the cut radius than the simulated sample, with a more elliptical core. 

\item {\it Weak Lensing, Bottom Right of Figure \ref{fig:differentComponentsRadius}}:
Finally, we find that the weak lensing shows a similar trend to that of the X-ray and BCG, with the two clusters MACJ1931 and MACSJ1206 showing a clear upward trend towards to the inner regions of the halo and significant outliers from the simulated sample. Moreover, three of the observed cluster sample are more elliptical than the entire simulated sample.
\end{itemize}

Following the measurement of the individual probes as function of radius we collate each distribution and show the estimated radial dependence of each probe in Figure \ref{fig:shapesDistance}. The top two panels in the left hand column shows the ellipticities of the observed clusters and the top two panels in the right hand column, the ellipticities of the simulated clusters.  The top row shows the absolute distributions (i.e. each top panel in the four figures of Figure \ref{fig:differentComponentsRadius}), with the addition of the estimates from strong lensing (in orange) and the moment of inertia derived from the particle data in black. The second row shows the estimated shapes relative to the strong lensing estimate. The third row shows the mis-alignment angle of each probe with the strong lensing estimate. 

In general we find that the distributions of observed ellipticities are consistent with one another, with no significant evidence for any radial dependent change in the ellipticity. Moreover, we find that the simulations and observations roughly agree with one another. Finally we find that the moment of inertia exhibits no radial dependence, with a constant ellipticity over the entire range.

The second row showing the results relative to the strong lensing shows that the observed ellipticities tend to be in agreement with the strong lensing estimate, apart from the BCG, which seems to be more elliptical for both the observed and simulated samples. Interestingly we find the X-ray and weak lensing estimates are both more spherical than the strong lensing, whilst the cluster members BCG and the moment of inertia are much more elliptical.   This is slightly inconsistent with \cite{clusterCompWithHorizon} who found that the probes all tended to be less elliptical than the strong lensing estimate.

The third row shows the alignment of each probe's major axis with respect to the strong lensing. The dashed line shows the 45 degree line representing to expected mean from random mis-alignment. We find for both the observed and simulated clusters all probes are aligned with the dark matter with a mis-alignment angle of $\sim 20$ degrees (depending on the cluster-centric radius). This is again consistent with  \cite{clusterCompWithHorizon} who found a mean mis-alignment angle of $\theta=22.2\pm3.9$ degrees. However, we do find that although the simulations are consistent with the observations, they do suggest a higher variance and random mis-alignments.

Finally it would be interesting to study the correlation of each probe with each other and other probes. As such, we carry out a complete test of the correlation of all probes and radial cuts. We define the correlation with the standard Pearson's correlation coefficient, 
\be
r_{xy} = \frac{ \sum ( x _i- \bar{x})(y_i-\bar{y})}{\sqrt{\sum(x - \bar{x})^2}\sqrt{\sum(y-\bar{y})^2}},
\ee
where barred quantities refer to the mean of the $x$ and $y$ samples. Figure \ref{fig:correlations} shows the correlation matrix for the ellipticity of each component and radial bin. We divide the matrix along the diagonal between observations (top left) and simulations (bottom right). We have used solid black lines to denote each probe (i.e. X-ray, strong lensing, weak lensing, BCG and cluster members, or `gals') and below the figure the colour-bar shows the correlation strength, with red showing an anti-correlation and green a positive one. We also correlate the moment of inertia, which is relevant only for the simulations, which is why this has the extra 5x5 matrix in the top left.
\figs
\includegraphics[width=\textwidth]{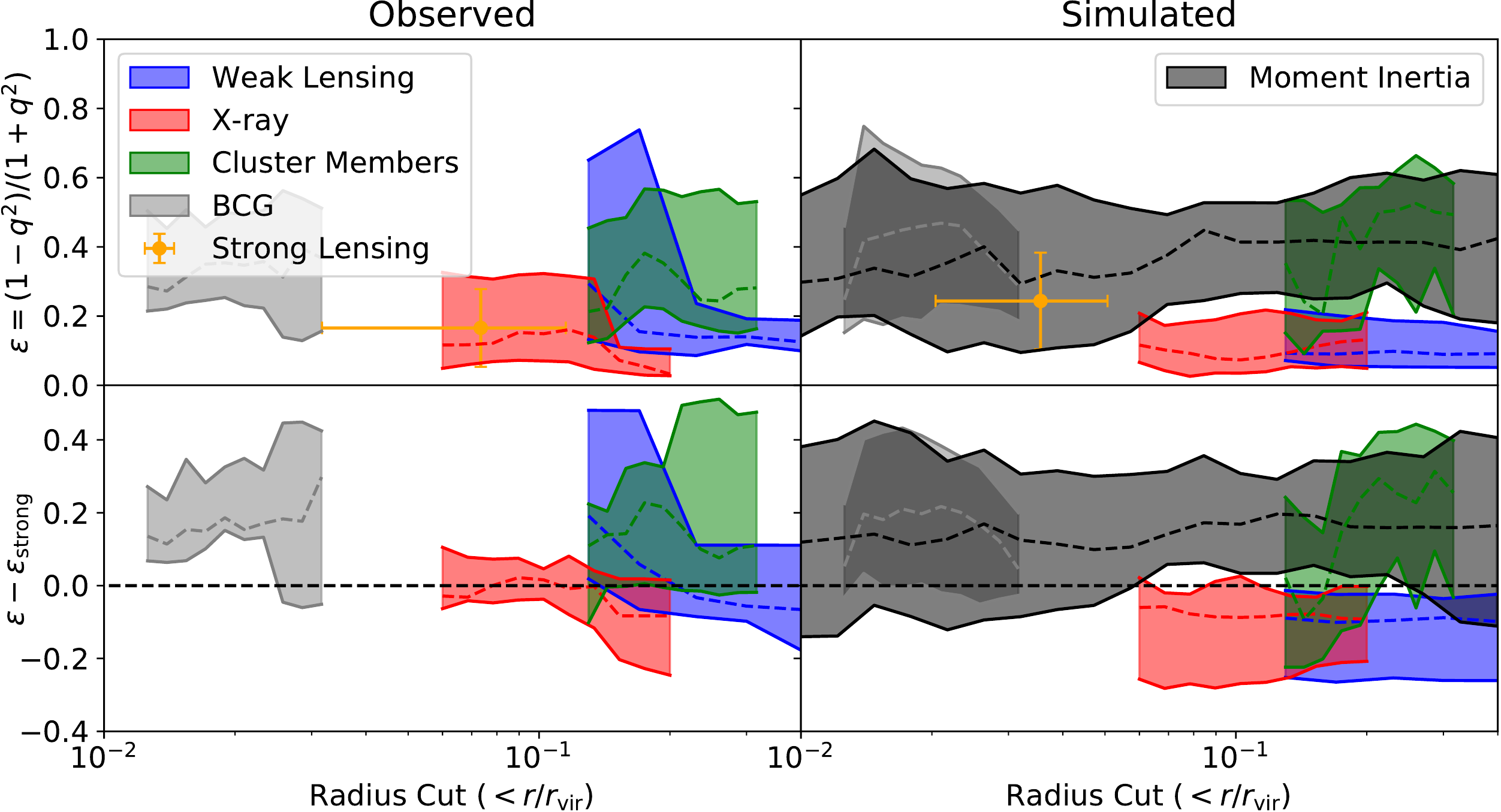}\\
\includegraphics[width=\textwidth]{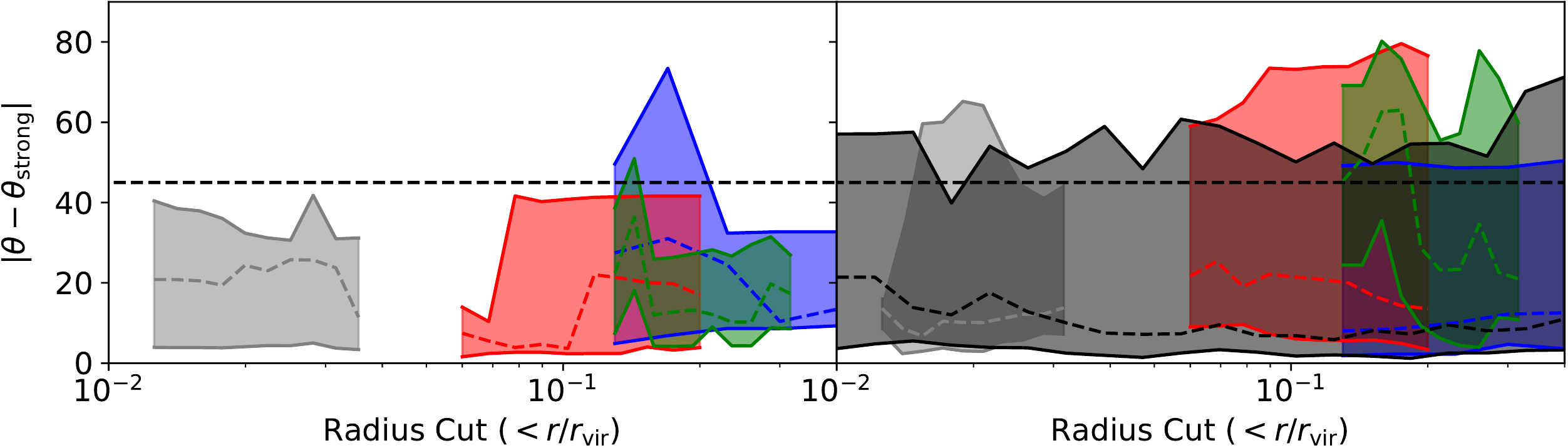}
		\caption{\label{fig:shapesDistance} The left (right) hand columns of the top two figures show the median, 84\% and 32\% distributions of the observed (simulated) ellipticity from the five different probes. The top row shows absolute values, with grey showing the shape of the BCG, the orange showing the strong lensing (with the width giving the range of virial radii this is measured at), the green shows the ellipticity derived from the cluster members, the blue gives the weak lensing shape, red gives the shape of the X-ray isophote and we show the moment of inertia derived from the particle data in black. The middle row shows the same probes, except relative to the strong lensing ellipticity. The bottom panel shows the mis-alignment of each cluster relative to the strong lensing ellipticity estimate. The dashed line at 45 degrees represents a random alignment.
 }
\efigs

The major difference between the two is that the cross-correlations between each component is stronger in the observations than the simulations. This could be due to the fact that  observed clusters are larger than the simulated ones and therefore any baryonic feedback in the simulations that could disrupt any correlation would have a larger (fractional) impact. 
\figs
\includegraphics[width=\textwidth]{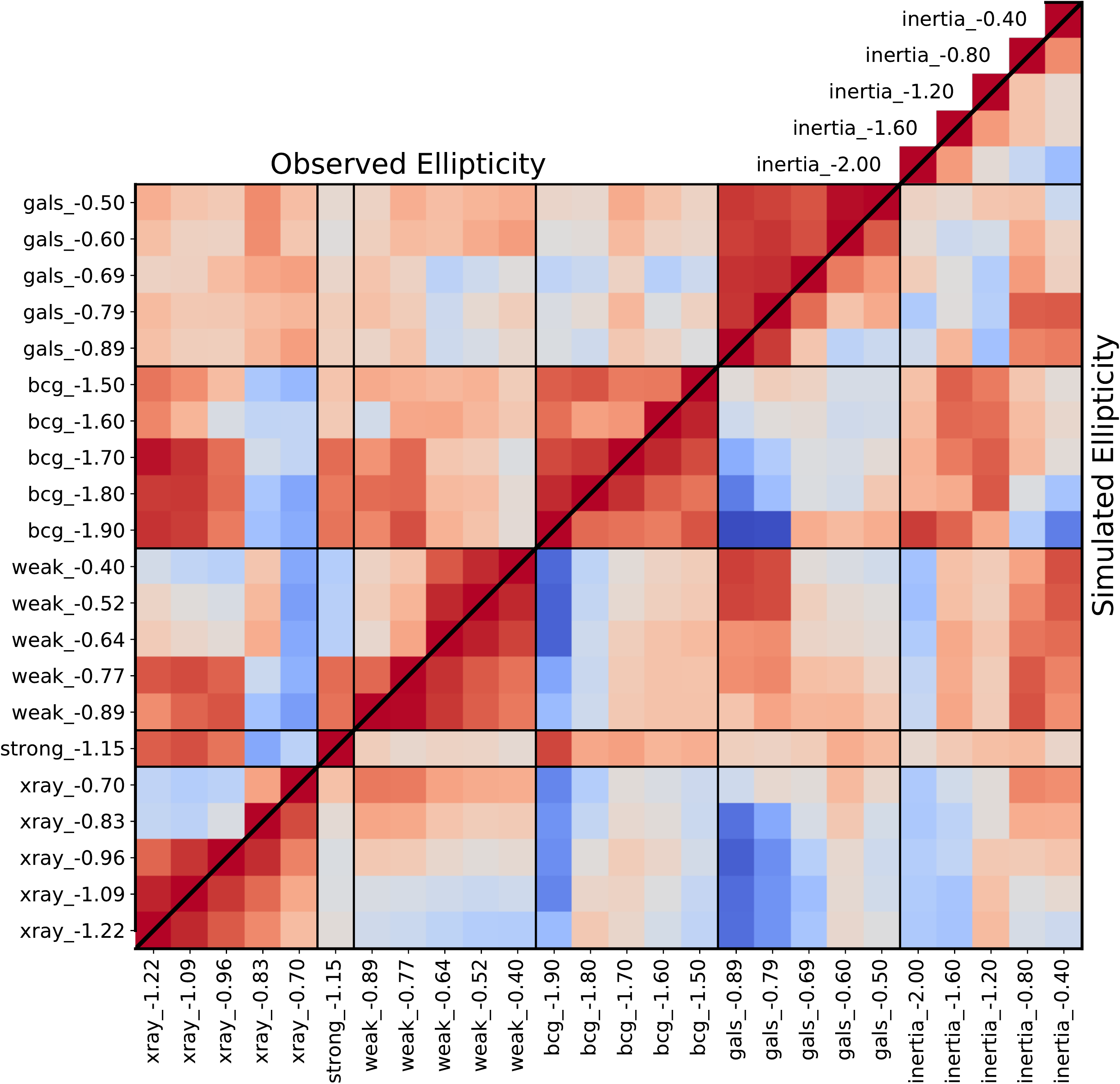}
\includegraphics[width=\textwidth]{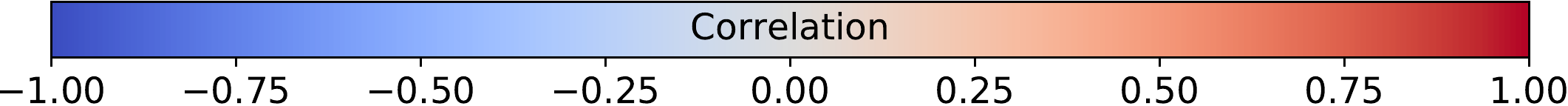}
\caption{\label{fig:correlations} The correlation between each probe and radial cut. The correlation is divided along the diagonal with the top left for the observed clusters and the bottom right for the simulated. Each label shows the type of probe and the log of the radial cut normalised to the virial radius of the cluster. The colour bar based at the bottom shows the correlation range, with red representing an anti-correlation and green a positive correlation. We note that `gals' corresponds to the shape derived from the distribution of cluster members and `inertia' the projected ellipticity derived from moment of inertia of the particle-data. We do not show the observed moment of inertia as this is not a directly observed quantity.
 }
\efigs
Now specifically looking at each probe, the key points to highlight are
\begin{itemize}
\item We find tentative evidence that the inner of regions of the observed X-ray halo do not correlate with the outer regions. This is mostly driven by the two cluster MACS1206 and MACS1931 and as such requires a greater sample to confirm, whereas the simulations show strong a correlation throughout. The cross-correlation with other probes shows that the inner regions of the X-ray isophotes correlate with the strong lensing and the smaller scales of the weak lensing. This is interesting if you compare to the simulated halos, which exhibit the opposite effect, with strong and weak both correlating better with the outer regions of the X-ray. This could be caused by the fact that the outer regions in the observed sample have a large amount of sky background, however the galaxy member shapes exhibit a  strong correlation with the X-ray shapes at theses scales, suggesting that this is not the case. Moreover, the simulated X-rays and galaxy distributions exhibit an anti-correlation. This could be due to the fact that the halos in the observed sample have a cooler core than the simulations, with less thermal motions disrupting the halo or a systematic of the lower of cluster members in these bins. We also find that outer regions of the X-ray are well correlated with the cluster members, something which we also observe in the simulated sample.
\item We find that the inner regions of the observed weak lensing correlate strongly with the strong lensing and the BCG, whereas the outer regions do not. This is intuitive since these are sensitive only to the inner regions.  We also find that the inner regions of the weak lensing correlate better with the inner regions of the cluster member shapes, however this is only  mild, whereas the simulated weak lensing exhibits as stronger correlation with the cluster member shapes, whilst a seemingly anti-correlation with the inner parts of the BCG.
\item The observed strong lensings shapes un-surprisingly have a  strong correlation with the BCG at all scales (including the pre-mentioned correlations with the inner regions of the weak lensing and the X-rays). They are also mildly correlated with the distribution of cluster members. A similar trend is found with the simulations, whereby the strong lensing correlates well with all probes.
\item Other than the already mentioned correlations of the BCG, (where its strongest correlation is with the inner regions of the clusters), we find that the outer regions of the observed BCG ellipticity is anti-correlated with the X-ray distribution. We do see signs of this anti-correlation with the inner regions of the simulated BCGs, with the outer regions of the BCG consistent with no correlation. What could drive this correlation is not clear and could be noise from the low ellipticity of the cluster members.
\item Finally we also correlate the two-dimensional moment of inertia (derived from the projected mass maps) with all simulated probes. We find that as expected the outer regions of the weak lensing and strong lensing do correlate well with inertia tensor. In addition, we find the outer regions of the X-ray isophote also agree with the moment of inertia. This is could be evidence that the inner regions of the hot gaseous halo is dominated by thermal motions, whereas the outer regions do tend to correlate with the shape of the cluster. We also find a mild correlation with the galaxy cluster members. The cluster members being tracers for the underlying halos should indeed have a correlation, however this is not as strong as we naively expected.
\end{itemize}

\section{Discussion and conclusions}\label{sec:conc}
We have carried out an investigation into the shape of eight dynamically {\it relaxed} galaxy clusters using a combination the {\it Hubble Space Telescope (HST)} and the {\it Chandra X-ray Observatory (CXO)} and compare them to the twenty-two most massive clusters in the BAHAMAS simulations in a bid to answer two key questions: {\bf 1.)} Is the ellipticity calculated from the projected moment of inertia derived directly from the particle data in simulations a good estimator of the shape derived from strong or weak lensing? {\bf 2.)} Is there any evidence for a radial dependent ellipticity in galaxy clusters, potentially signalling physics at different scales? 

To answer the first question we create mock strong and weak lensing observations from the BAHAMAS simulations and compare the ellipticity estimates to the projected moment of inertia calculated directly from the particle information. We find that in all cases the mock strong and weak lensing observations of the relaxed sample under-estimate the ellipticity with respect to the moment of inertia (at the same effective radius) by a factor$\langle\epsilon_{\rm S}/\epsilon_{\rm S, MI}\rangle= 0.64_{-0.04}^{+0.05}$ and $\langle\epsilon_{\rm W}/\epsilon_{\rm W, MI}\rangle=0.23_{-0.01}^{+0.01}$. respectively. In addition, we find that the position angle estimated from the mock weak lensing are  well aligned with the moment of inertia for both the relaxed and unrelaxed sample, yet the strong lensing is sensitive to substructures outside the critical curves, particularly for unrelaxed clusters, often inducing large mis-alignments. These results highlight the importance of deriving mock observations when comparing observations to simulations.

Following this we then use a combination of Brightest Cluster Galaxy, X-ray emission, distribution of galaxy members, weak gravitational lensing, and strong gravitational lensing, creating parallel mock observations in all cases to answer our second primary question. 

Studying the radial dependence of the ellipticity for each probe, we find that in general the broad distributions between the simulations and observations match well. The main discrepancies  we find is that the inner regions of MACSJ1206 and MACSJ1931 are much more elliptical than what is predicted by the simulations with the BCG, X-ray isophote and weak lensing observations all outliers to the expected distribution of the simulations.

 Indeed it is interesting that MACS1931 and MACS1206 show are discrepant  with respect to both the rest of the observed clusters and the simulated dataset. This is particularly intriguing since they exhibit no special X-ray luminosity. MACS1931 does have a particularly disturbed BCG which could suggest on-going feedback, however the BCG of MACS1206 is extremely regular. Having said this, \cite{MACS1206} found that despite MACS1206 being relaxed, there was significant asymmetry in the total mass profile, which again could suggest on-going feedback that could impact the shape. This question of the validity of the selection criteria when determining if a cluster is relaxed or merging and whether this needs to be more sophisticated than just the X-ray concentration, or (more importantly) it stresses the importance of matching selection criteria between simulated and observed samples since there are many micro-process in cluster that could be undetected, which have macro-impacts.

We study the auto and cross correlation between each probe and radial cut to understand how each probe relate to one another. We find in general all five mass components of the observed clusters trace the same underlying cluster shape, showing significant correlations between one another. In particular we find the ellipticity at similar radial cuts strongly correlate. For example the BCG ellipticity correlates with the strong lensing and the inner regions of the weak lensing, and the outer regions of the X-ray, galaxy members and weak lensing all correlate with one another. 

Interestingly, we do not find the same correlation with the simulated clusters, with trends differing from the observations. The small scales of each probe in these clusters exhibit much weaker correlations, with only the outer regions of the clusters showing the strong correlation seen in the observed clusters. We note that the cluster in the simulated sample are 0.5 dex less massive that the observed sample, which could be a cause of the weaker correlation.  In these smaller halos the fractional impact of feedback may be larger, disrupting the inner regions and weakening observed correlations. We also find that the cluster member distribution correlates with the X-ray. Finally we correlate the two-dimensional moment of inertia tensor derived from the projected particle data with each of the simulated probes and find that the weak and strong lensing strongly correlate and the X-ray emission in the outer regions also correlates. We find that the cluster member shapes, although tracers of the underlying structure, only  mildly correlates with the shape of the cluster from the projected moment of inertia. 

 We conclude that weak and strong lensing is a good {\it proxy} for the moment of inertia derived from the particle data, however they both significantly under-estimate the ellipticity. Therefore going forward if weak and strong lensing studies of the shapes of clusters are to be compared to simulations (in a bid to make statements on the nature of dark matter or impact of baryons in clusters), mock observations must be generated and analysed. 
 

Understanding the impact of baryons on massive clusters will be vital if we are to characterise how feedback alters the shape of clusters. Analysis directly comparing data to mock observations like this that probe different regions of the cluster will be important in this effort. Moreover, studies like these where exotic physics or modified gravity may change the shape of a cluster at all scales will also provide important tests of dark matter and general relativity.

%
%

This work has been presented in parallel with the public release of our shape measurement code {\sc pyRRG}. Available to directly install from PyPi via \url{https://pypi.org/project/pyRRG/}, this Python3.7 code based upon \cite{rrg} is specifically designed for HST shape measurement. It is fitted with an automated star-galaxy classifier and outputs scientifically useful products such as catalogues for the mass reconstruction code {\sc Lenstool}. For more see \url{https://github.com/davidharvey1986/pyRRG}.

\bsp

\section*{Acknowledgements}
DH acknowledges support by the ITP Delta foundation.
J.R. was supported by JPL, which is run under a contract for NASA by Caltech.
MJ is supported by the United Kingdom Research and Innovation (UKRI) Future Leaders Fellowship 'Using Cosmic Beasts to uncover the Nature of Dark Matter' [grant number MR/S017216/1]. This project was also supported by the Science and Technology Facilities Council [grant number ST/L00075X/1]. SIT is supported by Van Mildert College Trust PhD Scholarships. RM is supported by the Royal Society. AR is supported by the European Research Council's Horizon2020 project `EWC' (award AMD- 776247-6). This project has received funding from the European Research Council (ERC) under the European Union's Horizon 2020 research and innovation programme (grant agreement No 769130). 

\section*{Data Availability}
Although data is private, it is available upon request.

\label{lastpage}
\bibliographystyle{mn2e}
\bibliography{bibliography}
\appendix
\section{Shape measurement of different observational probes in data}\label{sec:shapes}

\subsection{The shape of the brightest cluster galaxy (BCG)}\label{sec:BCG}

The shape of the brightest cluster galaxy can be defined by its first order image moment,
\be
I=\int_0^R d^2\theta  i(\theta), ~~~~ \label{eqn:weightedMoment}
\ee
and its normalised quadrupole image moment,
\be
J_{ij}=I^{-1}\int_0^R d^2\theta~\theta_i\theta_j i(\theta),~~~~
\ee
where $i(\theta)$ is the flux at position $\theta$, $w(\theta)$. From this, the two components of ellipticity of a galaxy, $\chi_1$ and $\chi_2$, are 
\be
\chi_1 = \frac{J_{11} - J_{22}}{J_{11} + J_{22}},~~~~~~\chi_2 = \frac{2J_{21}}{J_{11} + J_{22}}, \label{eqn:ell}
\ee
where $\chi=(A^2-B^2)/(A^2+B^2)=\sqrt{(\chi_1^2+\chi_2^2)}$ and the size of the object, $d$, is given by the combination of the quadrupole moments,
\be
d=\sqrt{\frac{1}{2}(J_{11} + J_{22})}.
\ee
We then find that the error in $\chi_1$ and $\chi_2$,
\be
\label{eqn:ellBCG}
\sigma_{\chi_1}^2 = \frac{ \sigma_{xx}^2(1-\chi_1)^2 +  \sigma_{yy}^2(1+\chi_1)^2 - 2(1-\chi_1^2)\sigma_{xxyy}}{(J_{xx} + J_{yy})^2}
\ee
and
\be
\sigma_{\chi_2}^2 = \frac{ (\sigma_{xx}^2 +  \sigma_{yy}^2 + 2\sigma_{xxyy})\chi_2^2 +
	4(\sigma_{xy}^2-\chi_2(\sigma_{xxxy}*\sigma_{xyyy}))}{(J_{xx} + J_{yy})^2},
\ee
where $\sigma_{ij}$ is the error in the given second order moment,
\be
\sigma_{ij} = \frac{\int d^2\theta \sigma_i^2(\theta_i\theta_j - J_{ij})}{\int d^2\theta~i(\theta)},
\ee
and $\sigma_{ijkl}$ is the error in the fourth order moment, $J_{ijkl}$, where
\be
J_{ijkl}=I^{-1}\int d^2\theta~\theta_i\theta_j\theta_k\theta_l i(\theta),
\ee
and
\be
\sigma_{ijkl} = \frac{\int d^2\theta \sigma_i^2(\theta_i\theta_j - J_{ij})*(\theta_k\theta_l - J_{kl})}{\int d^2\theta~i(\theta)},
\ee
and the error in $i(\theta)$, $\sigma_i$, is,
\be
\sigma_i(\theta)^2 = \sigma_{\rm sky}^2 + \frac{i(\theta) - b(\theta)}{t_{\rm exp}},
\ee
where $\sigma_{\rm sky}$ is the estimated variance in the background and $b(\theta)$ is the estimated absolute background. To measure the radial dependence of the shape of the BCG we measure the total flux within some cut radius. To avoid bias due to the aperture when measuring the shape we carry out an iterative method whereby we calculate an initial ellipticity and then moderate the radial distances of each pixel to match the shape, such that
\be
r'=\sqrt{\theta_1^2 + \frac{\theta_2^2}{(1-\chi)}}.
\ee
We continue to iterate until the ellipticity converges to within a 1\% error. We limit the inner radius to a minimum of $r_{\rm cut}>20$kpc to avoid reaching the softening scale of the simulations ($4 h^{-1}$kpc) in both the simulations and the observations, ensuring an equal comparison.

\subsection{The shape of the distribution of cluster member galaxies}\label{sec:clusterMembers}

We follow a similar method to estimate the shape of the BCG whereby we measure the intensity weighted image moment, however with two key differences. To probe the radial dependence we measure the cumulative number of galaxies inside an cluster-centric cut radius R, and we measure the mass weighted moment similar to the moment of inertia tensor, such that
\be
I=\int_0^R d^2\theta M_*(\theta), ~~~~ \label{eqn:weightedMoment}
\ee
and its normalised, weighted moment quadrupole mass weighted moment,
\be
J_{ij}=I^{-1}\int_0^R d^2\theta~\theta_i\theta_j M_*(\theta),~~~~
\ee
where $M_*(\theta)$ is the total stellar mass and $R$ is the cluster-centric radius. We then calculate the two components of ellipticity using equation \eqref{eqn:ellBCG}. For the simulations $M_* = M_*(<100$kpc$)M_\odot$, for the observations where the complete spectral energy distribution (SED) is well sampled, i.e. the CLASH clusters \citep{CLASH_photoz}, we use LePhare to estimate the stellar mass of cluster members \citep{LePhare}.  LePhare is photometric redshift estimator that uses a SED fitting method. Here we fix the redshift to that of the cluster and constrain only the stellar mass and adopt a Calzetti extinction law \citep{calzettiExtinction}. We compare our stellar masses with publicly available catalogues from the CLASH website and find in general a tight correlation, however due to spurious outliers (stellar masses $>10^{14}$), in the public catalogue we decide to use the derived masses here. Where we do not have the spectral information for the cluster members (a1703 and a1835), we fit an empirical absolute magnitude (in the F814W band) - stellar mass relation and predict these. The top panel of Figure \ref{fig:stellarMasses} shows this relation. We see that the six CLASH clusters show a tight correlation and the black dashed line the fitted correlation. The bottom panel shows a histogram of the estimated stellar masses from the CLASH clusters. We see that they span below the mass-resolution of the simulations. Therefore to match the shapes in the simulations and observations we only estimate the shape from cluster members with $\log(M_*/M_\odot)>10$. In addition, since we are selecting cluster members from the red-sequence, we select only galaxies in the simulations that negligible star formation.

We derive error-bars empirically (since Poisson noise would prevent an accurate estimate using analytical estimates using the fourth order moments). To do this we measure the observed shape and then populate a field with the same ellipticity and angle and sample this randomly and re-measure the shape. We do this 100 times and find the upper and lower 34 percentiles around the observed estimate.

Finally we note that the mass of the clusters in the observed sample is larger than that of the simulated sample and as such we would expect there to be a difference the number density of cluster members. We hypothesise that the result would be a biased estimator of the shape of the distribution of cluster members in the event there are not many cluster members. As such we first measure the difference in the sub-halo mass functions of the two samples. The top panel of Figure \ref{fig:massFunct} shows how the number of cluster members in the observed sample is a factor of $\sim4$ greater than the simulated sample. Moreover, close to the core of the cluster the number of members is small. We test the impact this may have on estimation of the shape. To do this, we carry out a Monte Carlo simulations of a mock sample of cluster members. We distribute a sample of members assuming a Gaussian distribution (with a mean of zero and width of 10 arc-seconds, although this choice has no impact on the result), with an ellipticity of $\epsilon=0.4$ and re-calculate the ellipticity of the members for different sample sizes. The bottom panel of Figure \ref{fig:massFunct} shows that below 100 galaxies the shape becomes biased, however, our relatively large statistical error of $\sim0.1$, means that this does not become significant until $\sim10$ galaxies. We therefore limit our ellipticity measurements to bins that have at least 10 cluster members.

\fig
\includegraphics[width=0.5\textwidth]{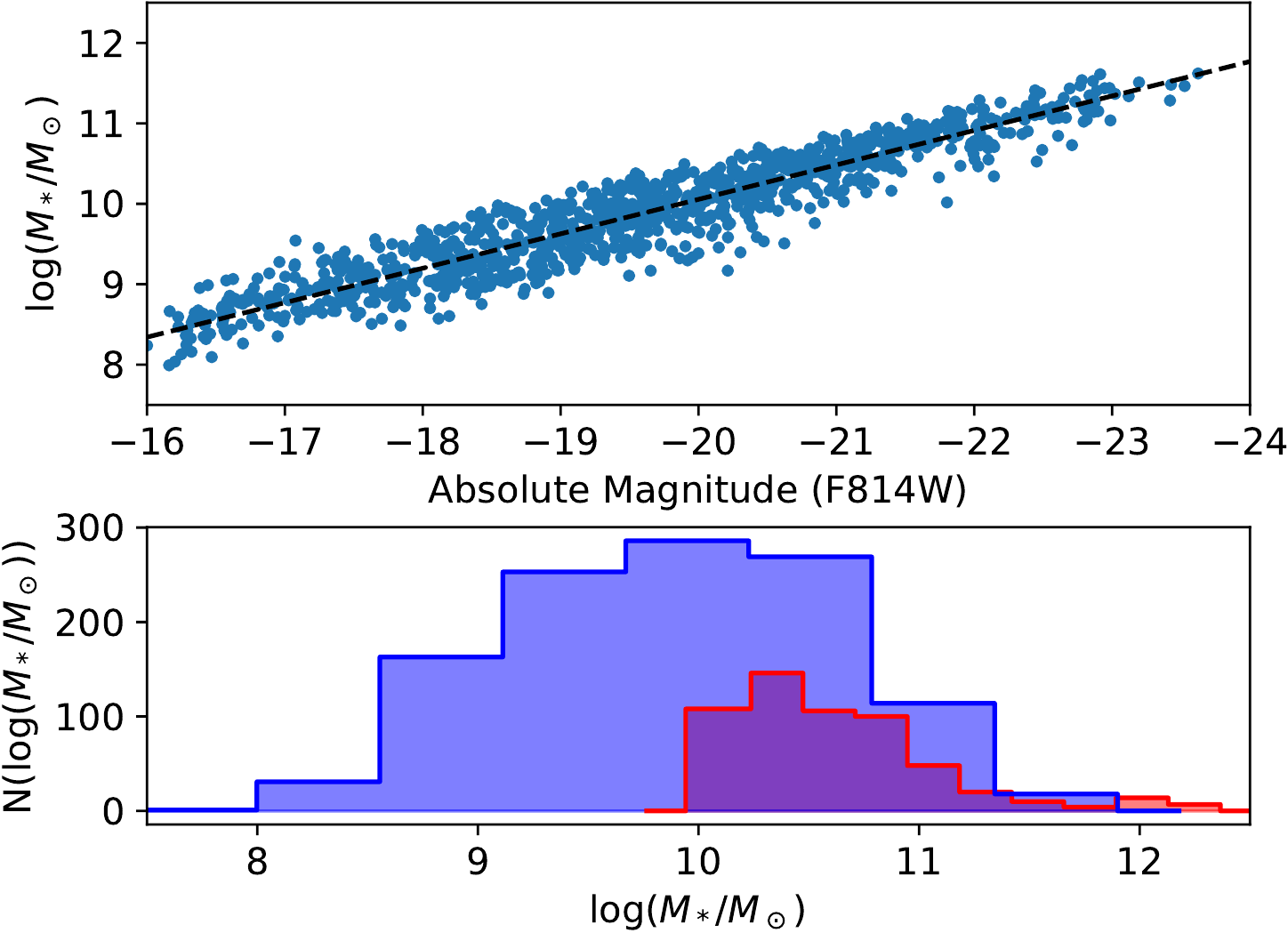}
\caption{\label{fig:stellarMasses} We estimate the stellar mass from the CLASH photometric catalogues \citep{CLASH_photoz} and the publicly available LePhare photometric-redshift estimator in order to estimate the shape of the distribution of cluster members using the moment of inertia tensor (top panel blue dots). For those clusters without high-fidelity photometric catalogues we fit an empirical relation between F814W absolute magnitude and estimate the stellar mass from their observed magnitudes (top panel dash line). The bottom panel shows the distribution of cluster member stellar masses from observations in blue and from simulations in red. We measure the moment of inertia tensor only from those cluster members with a mass greater than the mass threshold of the simulation ($\log(M_*/M_\odot)>10$) in order to match the two samples.}
\efig
\subsection{The shape of the Xray isophote}\label{sec:xray}

To measure the shape of the X-ray isophote we adopt the same technique as in the previous section for the galaxy shapes, however instead of weighting by the mass we weight by the exposure weighted flux. To measure the shape as a function of radial distance we follow the method laid out in \cite{shapeXrayClusters}. At each cumulative radial bin we iteratively estimate the the ellipticity by including all pixels inside the elliptical radius,
\be
r'=\sqrt{\theta_1^2 + \frac{\theta_2^2}{(1-\chi)}}.
\ee
At each iteration we estimate the elliptcity and angle and then re-estimate the shape including those pixels from the previous ellipticity estimate. We stop iterating when the ellipticity converges within an 1\% error. 

To estimate the error in the ellipticity we Monte Carlo each cluster image whereby we randomly re-sample each pixel assuming a Poisson distribution around the true value. We then re-estimate the shape of the cluster. We do this 100 times and take the 32\% percentiles around the observed value.

\fig
\includegraphics[width=0.49\textwidth]{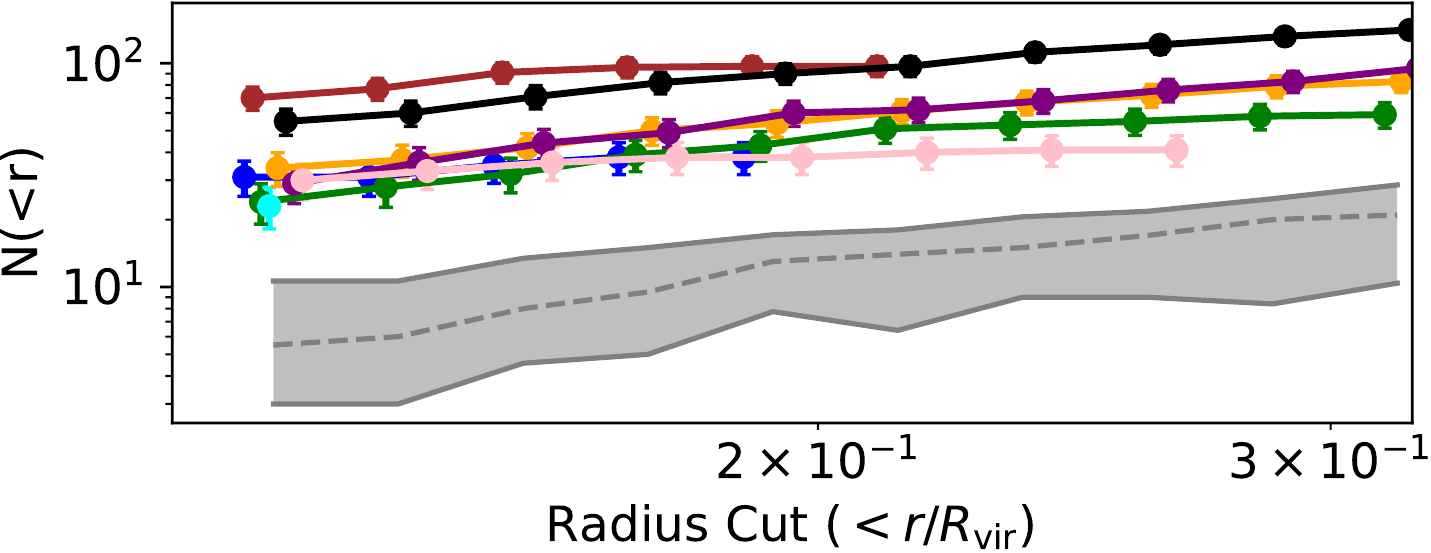}
\includegraphics[width=0.49\textwidth]{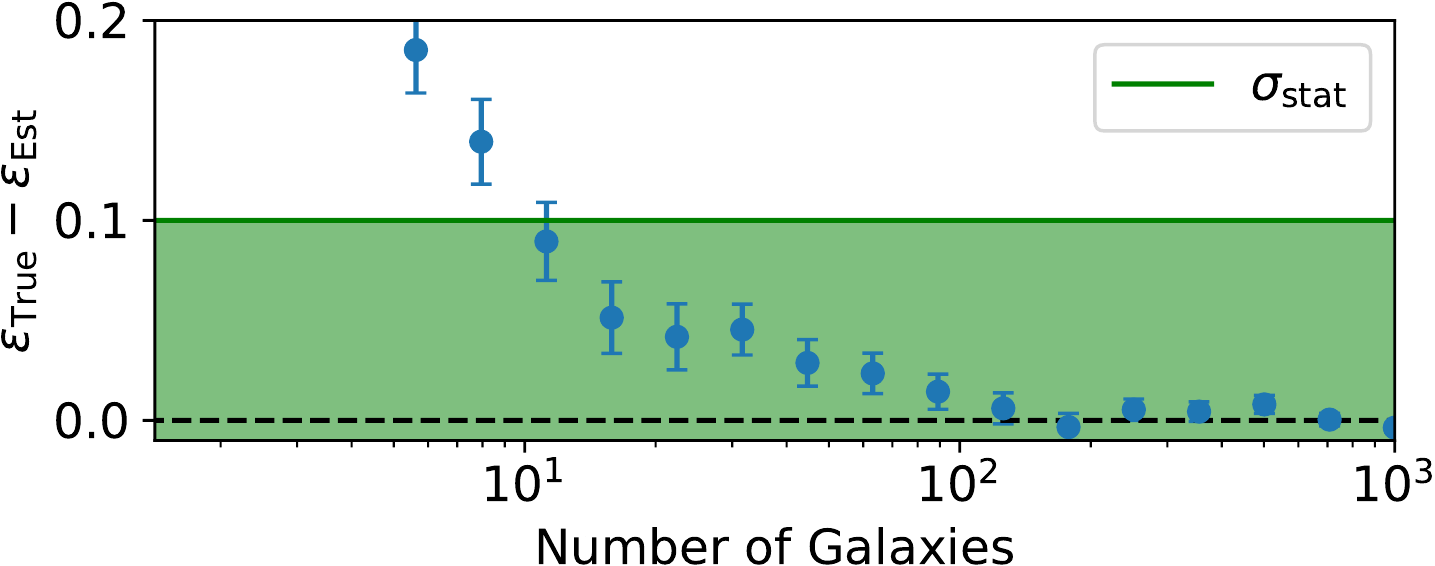}
\caption{\label{fig:massFunct} {\it Top Panel: } The radial distribution of sub-haloes for each cluster (coloured lines) and the simulations (grey region representing the 16\% and 84\% distribution around the median value (dashed line)). We find that the number of sub-halos is a factor of $\sim4$ greater in the observed sample that the simulated sample attributed to the difference in virial mass. {\it Bottom Panel: } The bias on the measured ellipticity of the cluster member distribution. We mock a simulated distribution of cluster members and re-estimate the ellipticity. We show the difference between the input and estimated as a function of the number of galaxy members available to measure the shape off. We find that below 10, the systematic bias becomes larger than the estimated, statistical bias of $\sigma_\epsilon = 0.1$ (green region). We therefore only select bins where there are $n>10$ galaxies.}
\efig

\subsection{The shape of the cluster from gravitational lensing}
We use the publicly available {\sc Lenstool} software which is a mass modelling algorithm that fits realistic parametric models to both strong and weak lensing observables to constrain the properties of the lensing potential \citep{lenstool}. It has become commonplace to use {\sc Lenstool} for combined strong and weak lensing analyses \citep[e.g.][]{MACSJ1149_HFF}, however here we want to analyse the difference in ellipticity between the core and the outer regions of the cluster. In this way we treat the strong and weak lensing reconstructions completely independently.

\subsubsection{Weak lensing mass mapping}\label{sec:weak}
To derive the weak lensing mass maps, we must first measure the weak lensing shear of all the background source galaxies in the cluster field. We do this via the publicly available code {\sc pyRRG} (see section \ref{sec:pyRRG}).
With the catalogues we use {\sc Lenstool} to estimate the weak lensing parameters. It does this by first projecting the observed ellipticities in the image plane to the source plane and then compares the subsequent source plane ellipticities with that expected from a Gaussian distribution with a width equal to the ellipticity dispersion of the sample, i.e.
\be
loss= \sum_{i=1}^2 \frac{ \chi_{i,s}^2}{\sigma_{\epsilon}},
\ee
where
\be
 \chi_{\rm s}= \frac{\chi_ - 2g + g^2\chi^\star}{1+|g|^2 -2\mathcal{R}(g\chi^\star)},
\ee
where we have assumed that the ellipticity can be written as a complex number in the form $\chi_1 = \chi + i\chi_2$, as produced from {\sc pyRRG}, the subscript 'S' denotes the ellipticity of the source and the star denotes the complex conjugate. We also adopt the mass and concentration from the strong lensing as a Gaussian prior on the weak lensing mass reconstruction.


\subsubsection{Strong lensing mass mapping}\label{sec:strong}
We follow the same procedure as used in \cite{Harvey_BCG} whereby we choose to model the global total matter halo with a Navarro, Frenk and White profile \citep{NFW} (and hence do not explicitly model the intra-cluster gas) and model each member galaxy, including the Brightest Cluster Galaxy (BCG) as a Pseudo Isothermal Elliptical Mass Distribution (PIEMD),
\be
\rho_{\rm NFW} \propto \frac{1}{x_{\rm NFW}(1+x_{\rm NFW})^2}
\ee
and
\be
\rho_{\rm PIEMD} \propto \frac{1}{(1+x_{\rm core}^2)(1+x_{\rm cut}^2)},
\ee
 where $x_{\rm NFW} = r / r_{\rm s}$, where $r_{\rm s}=r_{\rm vir}/c_{\rm vir}$,  $x_{\rm core} = r/r_{\rm core}$, $x_{\rm cut} = r/r_{\rm cut}$ and $r$ is the three dimensional cluster centric radius. Indeed, {\sc Lenstool} models the two dimensional NFW potential, $\Psi$ and not the mass distribution, $\Sigma$ and then assumes that $\epsilon_\Sigma = 3\epsilon_\Psi$. All ellipticities we report here are ellipticities of the mass distribution. For more we direct the reader to \cite{lenstool}. We also assume that the member galaxies fall on the fundamental plane (including the BCG) following a consistent mass to light ratio in order to reduce the parameter space such that for  the $i$th cluster member,
\be
r_{\rm core, i} = r_{\rm core}^\star\left(\frac{L}{L^\star}\right)^{1/2},
\ee
and
\be
r_{\rm cut, i} = r_{\rm cut}^\star\left(\frac{L}{L^\star}\right)^{1/2},
\ee
and that the velocity dispersion of the galaxy is 
\be
\sigma_i = \sigma^\star\left(\frac{L}{L^\star}\right)^{1/4}.
\ee
As is common amongst strong lensing reconstructions we assume $r_{\rm core}^\star = 0.15$kpc and we have a tight Gaussian prior of $\sigma^\star=158\pm26$km/s and $r_{\rm cut}^\star=45\pm1$kpc. Following this we have $6$ free NFW parameters from the main halo, and then two free parameters for the galaxy members. In rare cases we model individual galaxies as not doing this has shown to potentially bias mass reconstructions \citep{Harvey16}.

\subsubsection{Ellipticity error validation}
\cite{Harvey_BCG} found that the error estimate from {\sc Lenstool} on the Cartesian position of a dark matter halo using strong gravitational was underestimated by a minimum of an order of magnitude. In order to quantify whether or not the error reported using the width of the posterior distribution reasonably reflects the true error in the ellipticity we carry out two tests; one for the strong lensing observables and one for the weak lensing. 

To do this we follow the same method as \cite{Harvey_BCG}. We mock up twenty simulations based on the true data. Using the source positions from the data (for both weak and strong lensing), we use  the best fit mass model from the strong lensing reconstruction and project the sources to image positions to give a catalogue of weak and strong lensing image positions. In the case of the weak lensing we add noise through the random distribution of galaxy shapes, modelled by a Gaussian, with a mean of zero and a width equal to that of the true cluster. For the strong lensing we randomly shift the position, with the shift drawn from a Gaussian with a zero mean and a width of $0.5\arcsec$. We then reconstruct the mass distribution using {\sc Lenstool}. We Monte Carlo each cluster $100$ times for the weak lensing and $10$ for the strong, since the strong lensing reconstructions take significantly longer to converge.

We find that in the weak lensing case, the broad posterior is biased when ellipticity is sampled in polar coordinates, as such we sample the data in Cartesian coordinates; $\epsilon_1$ and $\epsilon_2$, where we defined $\epsilon=(a^2-b^2)/(a^2+b^2)$ and $a$ and $b$ are the semi major and minor axes of the mass distribution respectively. We then convert the results back to polar coordinates to be consistent with other probes. Figure \ref{fig:strongPosteriorWeakMonteCarloComp} show the results of the strong and weak lensing reconstructions. Each panel shows the estimate of the variance in the Monte Carlo tests as a function of the direct estimate of the error from the width of the posterior sampled during the MCMC of the true data. The left (right) hand column shows the results from the strong (weak) lensing. We find that the estimate of the error from the posterior in the strong lensing significantly under-estimates the true error from the Monte Carlo tests, whereas the weak-lensing the posteriors slightly over estimate the error. As such during the analysis we take the error in the strong lensing from the variance in the Monte Carlo tests, and the weak lensing directly from the posterior.

\subsection{Shape from the moment of inertia using simulated particle data}\label{sec:MI}

Unlike some of our observational shape definitions, our moment of inertia calculation does not iteratively fit for the centre of the halo. Instead, the 2D centre of the halo is defined as the location (in projection) of the particle with the most negative gravitational potential energy, and this is kept fixed throughout the calculation. The moment of inertia calculation is done in 2D, having first projected all particles within $5 \, r_{200}$ of the cluster centre (in 3D) along the relevant line-of-sight. For a given radius, $r_\mathrm{cut}$, we begin by finding all particles in a circle of radius $r_\mathrm{cut}$. The inertia tensor
\begin{equation}
I_{ij} \equiv \sum_n x_{i,n} \, x_{j,n} \, m_{n} \, \Big/  \, \sum_n m_{n}
\label{mass_tensor}
\end{equation}
is calculated for this distribution of particles, where $(x_{1,n},x_{2,n})$ are the coordinates of the $n$th particle, which has mass $m_{n}$.  We label the eigenvalues of $I_{ij}$ as $a^2$ and $b^2$ (with $a \geq b$), and the corresponding eigenvectors as $\vect{e_1}$ and $\vect{e_2}$. The axis ratio is $q=b/a$.

Our process is iterative, and in each iteration we calculate the inertia tensor for the particles within an ellipse, where the axis ratio of the ellipse is determined by $q$ from the previous iteration, and the major axis of the ellipse is along the $\vect{e_1}$ direction. The area of the ellipse is kept constant, meaning that it has a semi-major axis of length $r_\mathrm{cut} / \sqrt{q}$ and a semi-minor axis of length $r_\mathrm{cut} \sqrt{q}$. We continue this iterative process until consecutive iterations agree on $q$ to better than 1\%.

\section{Shape Measurement Algorithm: PYRRG}\label{sec:pyRRG}
\fig
\includegraphics[width=0.5\textwidth]{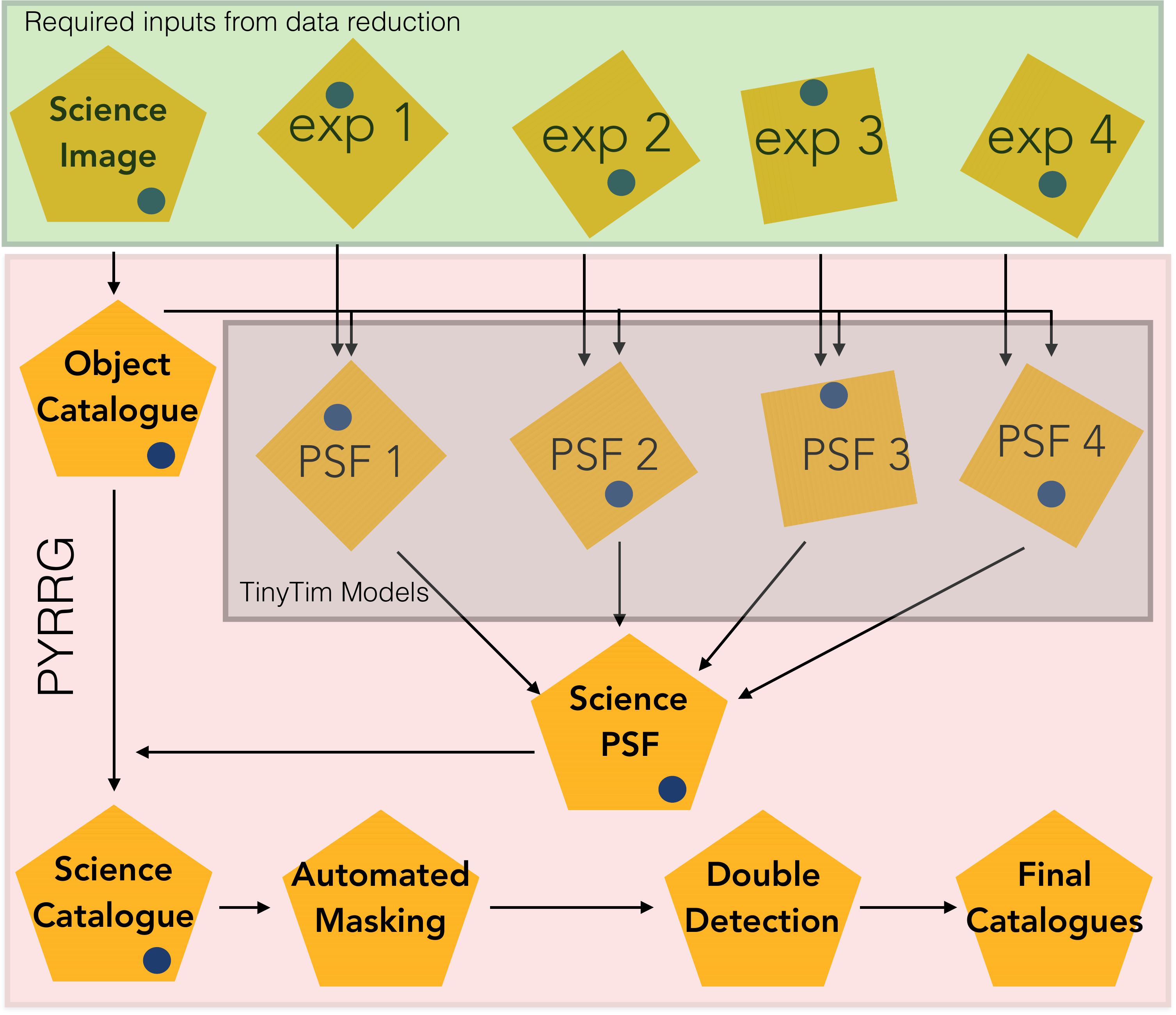}
\caption{\label{fig:PYRRG} An overview of the {\sc pyRRG} algorithm. It requires the input science image and associated weight file from the data reduction pipeline, plus all the associated exposures. From this it chooses the `best' PSF from the TinyTim models and combines them to produce a PSF at the position of each galaxy in the catalogue. It then corrects the galaxies, calculate the shears and outputs a science catalogue. It then carries out post-processing procedures to create a final `clean' catalogue.}
\efig
In this section we briefly outline the publicly available shape measurement software, specifically designed for images from the Advanced Camera for Surveys on the Hubble Space Telescope. We begin by first outlining the theoretical basics.  For a review please see \cite{BS01}, \cite{MKRev}, \cite{HoekstraRev} and \cite{RefregierRev}. For a review on strong and weak gravitational lensing please see \citep{gravitational_lensing}. 

Gravitational lensing is simply a distortion of a background source at a position, $\beta$ by foreground matter, inducing a shift of $\hat{\alpha}$ to the observed position $\theta$, i.e. 
\be
\beta = \theta - \frac{D_{\rm LS}}{D_{\rm S}}\hat{\alpha} = \theta - \alpha,
\ee
where we have introduced the reduced  deflection angle, $\alpha$. This deflection angle is related to the potential causing the deflection, which is the projected three dimensional Newtonian potential, $\Phi$,
\be
 \nabla \Psi = \alpha,
 \ee
 where 
 \be
 \Psi = \frac{D_{\rm s}}{D_{\rm L}D_{\rm S}}\frac{2}{c^2}\int\Phi(D_{\rm L},\theta,z) dz.
 \ee
 From this we can derive the distortion matrix by examining how a change in the source position effects the change in the image position, i.e. ${\partial \theta}/{\partial \beta}$, also known as the lensing Jacobian,
 \be
 A_{ij} = \frac{\partial \beta}{\partial \theta} = \delta_{ij} - \frac{\partial^2 \Psi}{\partial \theta_i\theta_j} = \delta_{ij} - \Psi_{ij},
 \ee
 where we have denoted the second derivative of $\Psi$ by the subscript, $i$ and $j$.
The second derivative of the lensing potential gives the two observables, the convergence, which is the trace of $A$,
\be
\kappa = \frac{1}{2}(\Psi_{11} + \Psi_{22}),
\ee
and corresponds to a scalar increase or decrease in the size of a distorted background source and the shear, $\gamma$ is a two component vector field given by,
\be
\gamma_1 = \frac{1}{2}(\Psi_{11} - \Psi_{22})~~~~{\rm and }~~~~ \gamma_2 = \Psi_{12} = \Psi_{21},
\ee 
corresponding to a stretch along the x-axis for $\gamma_1$ and $45^\circ$ for $\gamma_2$. We now have a relation between the observable distortion and the lensing potential. Here we limit the expansion of the Jacobian to first order, and hence assume a weak lensing limit. Indeed the shear and convergence are coupled and one cannot be observed without the other, this is known as the reduced shear,
\be
g=\gamma / (1 - \kappa).
\ee

\subsection{Shape Measurement: {\sc pyRRG}}
The weak lensing shape measurement consists of six key sections. An overview of the {\sc pyRRG} algorithm can be found in Figure \ref{fig:PYRRG}.
\begin{enumerate}
\item Source finding
\item Moment measuring
\item Star-Galaxy Classification
\item Point Spread Function estimation
\item Shear estimation
\item Catalogue cleaning \& masking
\end{enumerate}

\subsection{Source finding}
{\sc pyRRG} employs the `hot and cold' method that was originally developed in \cite{COSMOSintdisp} to extract sources from the image and then extended to studies used in \cite{cosmicBeast,MACSJ1149_HFF,MACSJ0717_HFF,A2744_HFF} and \cite{Harvey15_quasars}. Using the open source program {\sc SExtractor} \citep{sextractor} {\sc pyRRG} carries out two scans of the image. The first, `hot' scan, uses a smaller minimum number of pixels to count as a source, thus finding smaller objects. The second, `cold' scan, uses a larger number of pixels to classify a source. We then use the publicly available {\sc Stilts}\footnote{\url{http://www.star.bris.ac.uk/~mbt/stilts/}} software to combine the two catalogues in to one final catalogue.

\subsection{Moment measurement}
Following the source detection, {\sc pyRRG} measures the weighted multipole moments of each object in order to characterise the shape. For a full description please see \cite{rrg}, however here we outline the basics. We define the zeroth order multiple moment of a two-dimensional image in $\theta$, with an intensity distribution $i$,
\be
I=\int d^2\theta w(\theta) i(\theta), ~~~~ \label{eqn:weightedMoment}
\ee
then the quadruple {\it normalised}, weighted moment is,
\be
J_{ij}=I^{-1}\int d^2\theta~\theta_i\theta_j w(\theta) i(\theta),~~~~
\ee
followed by the fourth order,
\be
J_{ijk}=I^{-1}\int d^2\theta~\theta_i\theta_j\theta_k w(\theta) i(\theta),
\ee
 where the weight is simply a Gaussian with a width $w$, where $w=\sqrt{(A_{\rm SEX}B_{\rm SEX}/\pi)}$, AND $A$ and $B$ are the semi major and semi minor axes as estimated by {\sc Sextractor}.
From this we can define the two components of ellipticity of a galaxy, $\chi_1$ and $\chi_2$, as 
\be
\chi_1 = \frac{J_{11} - J_{22}}{J_{11} + J_{22}},~~~~~~\chi_2 = \frac{2J_{21}}{J_{11} + J_{22}}, \label{eqn:ell}
\ee
and the size of the object, $d$, is given by the combination of the quadrupole moments,
\be
d=\sqrt{\frac{1}{2}(J_{11} + J_{22})}.
\ee

\subsection{Star - Galaxy Classification: Random Forest}

Following the measurement of the normalised image moments, we classify objects in to three distinct categories, stars (both saturated and not), galaxies and noise. Given that it is a simple classifying problem, we adopt a Random Forest to automatically classify this.

A Random Forest is a supervised machine learning tool that generates an ensemble of decision trees that are then trained on known data to produce predictions for unknown data \citep{randomForest}. It generates a single tree by randomly subsampling the data and carrying out simple regression to create an estimator for the subsample of data. It then re-samples randomly with replacement to generate another tree. Given that each tree is a poor unbiased estimator of the truth, the aggregated estimator should be the correct one. 
The number of trees defines how good the overall estimator is, but also how long it takes to train and how large the classifier is. 

We generate a range of data to train the Random Forest. We use data from {\it HST} including a sample of $21$ SLACS galaxies, $29$ galaxy clusters, all at a range of depths. This way we try to span the entire range of parameter space including, object magnitudes, environment, and signal to noise. We generate the ground truth by manually classifying stars and galaxies from their magnitudes, $\mu_{\rm max}$, and size for individual exposures.
We then aggregate this data and parse through the Random Forest all information regarding these objects including the object magnitude, the size of the object, the brightest pixel in the object ($\mu_{\rm max}$), the second and fourth order moments, the uncorrected ellipticities, the median sky background and the variance around this, and finally the exposure time of the image. The main panel of Figure \ref{fig:classifierImportance} shows the relative importance of each feature in the classifier, where relative importance is a unit-less coefficient that defines the importance of each feature in discriminating between classifiers, i.e. when the data is subsampled, each decision tree will define the most important classifier for that sub-sample. Over the entire ensemble of decision trees, this importance reflects at which level in the decision tree this feature lies. The inset shows the result on a blind test galaxy, GAL-0364-52000-084. We find that the Random Forest has a $93$\% rate of correctly classifying stars, $99$\% rate of classifying galaxies and $83$\% of characterising noise (including saturated stars)\footnote{{\sc pyRRG} allows manual selection of galaxies through an interactive region selecting scheme, however the default and what is used for this work is the Random Forest}.

\fig
\includegraphics[width=0.5\textwidth]{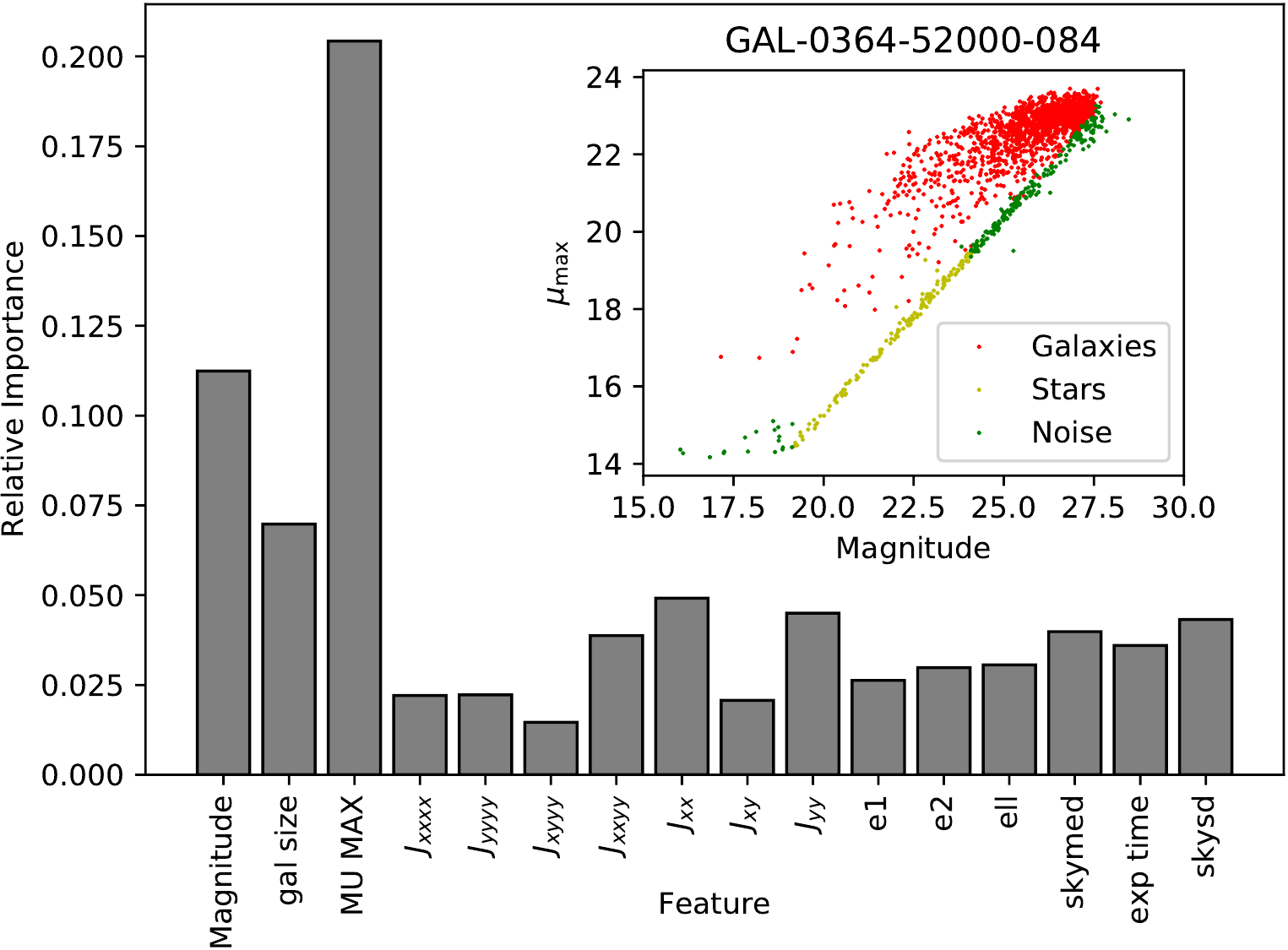}
\caption{\label{fig:classifierImportance} Results of the automated star-galaxy classifier. We use a Random Forest trained on data from the {\it HST} to predict the classification of each object. The main panel shows the relative importance of each feature in the classifier, the inset shows the results of classifying objects for the object {\it HST} field with GAL-0364-52000-084. We find that the Random Forest has a $93$\% rate of correctly classifying stars, $99$\% rate of classifying galaxies and $83$\% of characterising noise (including saturated stars).}
\efig

\subsection{Point Spread Function Measurement}

Having classified the stars, {\sc pyRRG} then estimates the impact of the telescope on the image, i.e. the Point Spread Function (PSF). HST warms up and cools down due to the heating of the Sun and therefore the focus of the telescope changes over time. We therefore estimate the impact of this by 
\begin{enumerate}
\item Taking the known positions of stars from the drizzled science image and finding the corresponding position in the individual exposures that make up that image.
\item Measuring the second and fourth order moments of the stars in each of the {\it individual exposures}
\item Having measured the moments, we compare to the various Tiny Tim models of the PSF \citep{tinytim}. We then interpolate this model to the known positions of the galaxies.
\item Combining the PSFs from each individual exposure at the position of the galaxy by rotating each PSF moment through an angle $\phi$ to the same reference frame as the drizzled science image in order to find the new rotated moment, $J'$, \citep{rotatemoments},
\begin{multline}
J'_{jk} = \sum^j_{r=0}\sum^k_{s=0}(-1)^{k-s}{j\choose k}{k\choose s}\\
\times (\cos\phi)^{j-r+s}
(\sin\phi)^{k+r-s}(J_{j+k-r-s,r+s}),
\end{multline}
and then summing the moments for a given position of the galaxy to acquire the final PSF.
\end{enumerate}
\subsection{Shear estimation}
Following the estimation of the PSF we then correct the galaxy moments and calculate the shear. 
As shown in \cite{rrg}, the final estimated shear is given by 
\be
\gamma_i = \langle\chi_i\rangle / G,
\ee
where $\chi_i$ is given by equation \eqref{eqn:ell} and with
\be
G=2-\langle\chi^2\rangle - \frac{1}{2}\langle\lambda\rangle- \frac{1}{2}\langle\chi\cdot\mu\rangle,
\ee
where $\langle\chi\cdot\mu\rangle = \chi_1\mu_1 + \chi_2\mu_2$,
\be
\lambda = (J_{1111}+2J_{1122}+J_{2222})/(2d^2w^2),
\ee
and the two components of the spinor, $\mu$,
\begin{align*} 
\mu_1 &= (-J_{1111}+J_{2222})/(2d^2w^2), \\
\mu_2 &=-2(J_{1112} + J_{1222})/(2d^2w^2), \addtocounter{equation}{1}\tag{\theequation}
\end{align*}
where $w$ is the size of the weight function $w(\theta)$ in equation \eqref{eqn:weightedMoment}. From this we have a final estimator of the shear, $\gamma$.
\fig
\includegraphics[width=0.5\textwidth]{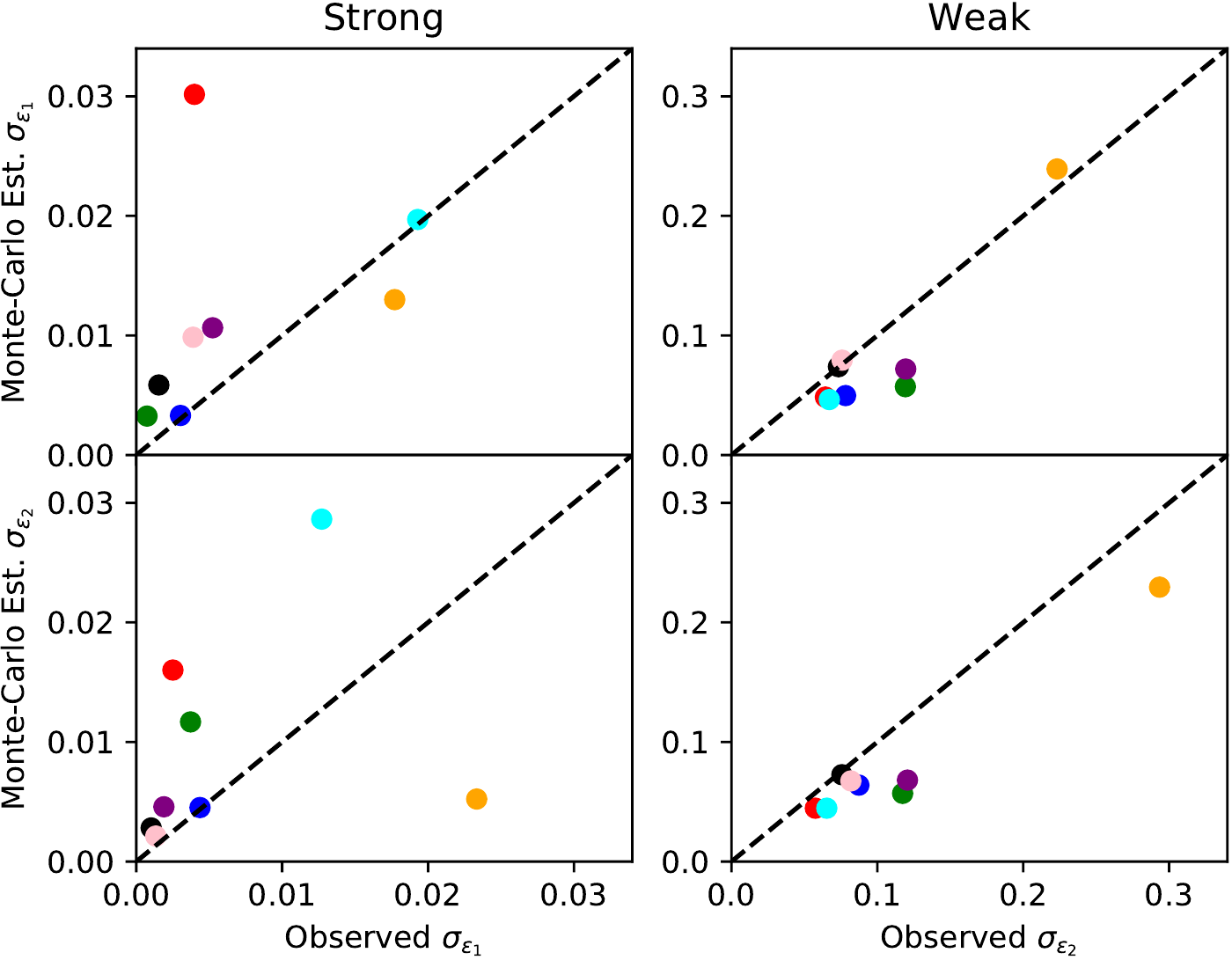}
\caption{Gravitational lensing error validation. We Monte Carlo the best fit strong (left column) and weak (right column) lensing model 20 and 100  times respectively. We show the error from the estimated posterior on the x-axis and the variance in the Monte-Carlo on the y-axis for $chi_1$ in the top panel and $\chi_2$ in the bottom panel.\label{fig:strongPosteriorWeakMonteCarloComp}}
\efig
\subsection{ Catalogue cleaning \& masking}\label{sec:postProcess}
Having measured the shear we go through a series of cleaning operations including,
\begin{itemize}
\item{\it Automatic Masking: } Using the known position of stars and saturated stars, we generate polygons that have the same size as stars and mask any object that lies within these polygons,
\item{\it Removal of double detections: } We remove double detections whereby removing objects that lie within the isophote of a larger object.
\item{\it Creation of a {\sc Lenstool} catalogue: } Creates a catalogue to be parsed in to the mass mapping algorithm {\sc lenstool}.
\end{itemize}
Finally we match the catalogues with the CLASH catalogues, which have accurate photometric redshifts \citep{CLASH_photoz} in order to remove contaminations. This includes cluster members and sources that are in the strong lensing regime (and hence cannot be used for the weak lensing estimates).

We first remove all galaxies that lie in the cluster. To do this we remove all galaxies that lie at $z>z_{\rm cl}+ \delta z$. The value of $\delta z$ depends on the cluster, \cite{redshiftCut} used a value of $\delta z = 0.2$. Here we test the impact of this cut on our sample and cluster shapes. Figure \ref{fig:testDeltaZ} shows the estimated $\epsilon_1$ ($\epsilon_2$) shown in the top (bottom) panel relative to the estimated shape with a cut of $\delta z=0.05$. We find that in general the clusters are not sensitive to this cut except MACSJ1206, which exhibits significant sensitivity. We find that a small cluster of galaxies at $z\sim0.7$ is pushing the fit to negative $\epsilon_1$. We therefore select a cut of  $\delta z=0.05$ for clusters except MACSJ1206 where we choose  $\delta z=0.35$.
We then remove all source galaxies that are within the strong lensing region. \cite{weakLensingInStrongRegion} found that moment based methods such as RRG are accurate to within $7$\% beyond the Einstein Radius of the cluster at which point by galaxies being to be strongly lensed and weak lensing assumptions break down. We therefore conservatively extract all galaxies up to 10\% beyond the estimated Einstein Radius of the cluster to ensure we are not including arcs and flexed galaxies.

\fig
\includegraphics[width=0.5\textwidth]{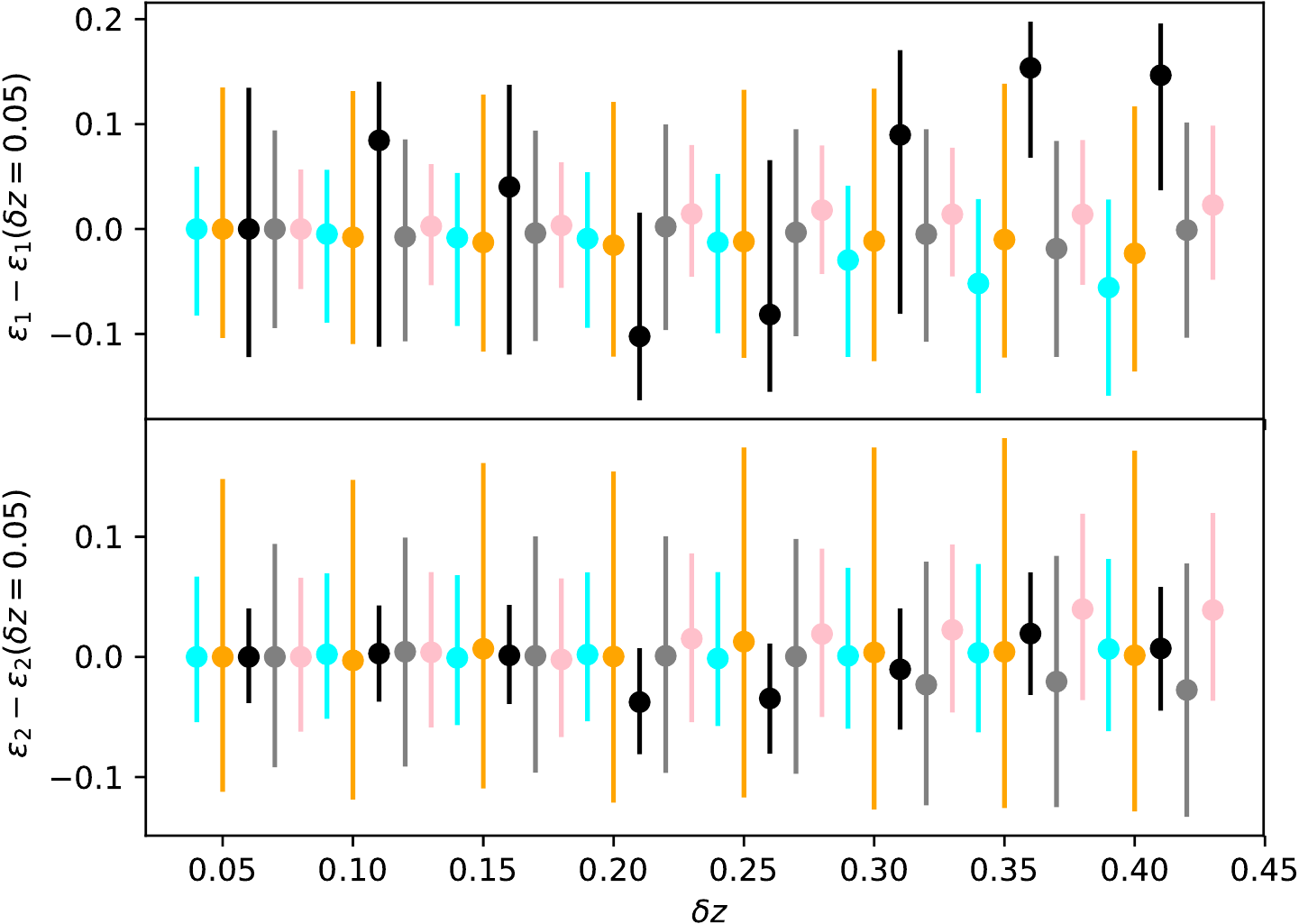}
\caption{\label{fig:testDeltaZ} The dependence of estimate of the cluster shape from weak lensing on the source galaxy redshift cut. We cut the galaxy such that only those galaxies in the catalog are $z_s > z_{\rm cluster} +\delta z$. The top panel shows the first component of ellipticity $\epsilon_1$ and the bottom panel $\epsilon_2$. We show for each cluster (different colours) with respect to the shape estimate for $\delta z=0.05$. We offset the clusters at each cut value for clarity. We find that the only cluster that is sensitive to this choice is MACSJ1206, where a small cluster in redshift space at $\sim0.7$ drives the fit. }
\efig

\end{document}